\documentclass[twocolumn,showpacs,preprintnumbers,amsmath,amssymb]{revtex4}
\usepackage{graphicx}
\usepackage{dcolumn}
\usepackage{bm}
\usepackage{epstopdf}
\usepackage[usenames,dvipsnames]{color}

\begin{document}

\title{Single Equalizer Strategy with no Information Transfer for Conflict Escalation}

\author{A. Engel}
\affiliation{Physics Department, Bar Ilan University, Ramat Gan, Israel}

\author{A. Feigel}
\email{sasha@phys.huji.ac.il}
\affiliation{Racah Inst. of Physics, Hebrew University of Jerusalem, Israel}

\date{\today}

\begin{abstract}
In an iterated two-person game, for instance prisoner’s dilemma or the snowdrift game, there exist strategies that force the payoffs of the opponents to be equal. These equalizer strategies form a subset of the more general zero-determinant strategies that unilaterally set the payoff of an opponent. A challenge in the attempts to understand the role of these strategies in the evolution of animal behavior is the lack of iterations in the fights for mating opportunities or territory control. We show that an arbitrary two-parameter strategy may possess a corresponding equalizer strategy which produces the same result: statistics of the fight outcomes in the contests with mutants are the same for each of these two strategies. Therefore, analyzing only the equalizer strategy space may be sufficient to predict animal behavior if nature, indeed, reduces (marginalizes) complex strategies to equalizer strategy space. The work's main finding is that there is a unique equalizer strategy that predicts fight outcomes without mutual cooperation. The lack of mutual cooperation is a common trait in conflict escalation contests that generally require a clear winner. In addition this unique strategy does not assess information of the opponent's state. The method bypasses the standard analysis of evolutionary stability. The results fit well the observations of combat between male bowl and doily spiders and support an empirical assumption of the war of attrition model that the species use only information regarding their own state during conflict escalation.
\end{abstract}

\maketitle

\section{Introduction}
\label{sec:org7ab72fa}

Animals fight for mating opportunities and territory control\cite{Hardy2013}. The contest, generally, proceeds as a continuous fight or a series of aggressive encounters until one of the competitors either dies or flees. The winner gains a resource that contributes to its Darwinian fitness. The field studies of animal combats pose a theoretical challenge to predict observed strategies for fights as a function of the observed evolutionary payoffs\cite{Milinski2014}. For instance, what are the payoffs that cause the spices to assess (mutual assessment) or to discard (self assessment) information regarding their opponent during the fight\cite{Taylor2003,Arnott2009,Briffa2009,Fawcett2013,Mesterton-Gibbons2014}?

Two-person games, e.g. prisoner's dilemma or the snowdrift game, constitute a well-established framework for analyzing evolutionarily optimal behavior\cite{Smith1982,AXELROD1981,Sigmund1998,Dugatkin2000,Nowak2006,Stewart2016}. In a two-person game, players adopt either of the two distinct behavior roles. We will use a defector \(D\) and a cooperator \(C\) as standard notations of the roles. There are four possible outcomes for a single round of the game between players \(h\) (host) and \(m\) (mutant): \((CC,CD,DC,DD)\) where both players cooperate, \(h\) cooperates while \(m\) defects, \(h\) defects while \(m\) cooperates, and where both players defect. The payoffs of player \(h\) are: 
\begin{eqnarray}
\label{eq:1}
 \vec{W}(h)=(R,S,T,P), 
\end{eqnarray}
for the outcomes \((CC,CD,DC,DD)\), respectively.

The type of game depends on its payoffs (\ref{eq:1})\cite{Sigmund2010}. Prisoner's dilemma \(T>R>P>S\) severely punishes cooperation \(C\) with a defector \(D\): the minimal payoff is \(S\) for the outcome \(CD\). In conflict escalation, e.g. the snowdrift games \(T>R>S>P\), mutual defection \(DD\) claims a heavy price from both sides similar to a competition of two cars that race toward each other. In this article, conflict escalation is defined as \(R>S>P,T>S\). This definition includes the snowdrift game (which favors defection against a cooperator \(DC\)) and a similar game that may favor mutual cooperation \(CC\). 

The gain \(G(h,m)\) of a player \(h\) against \(m\) is: 
  \begin{eqnarray}
\label{eq:44}
G(h,m)=\vec{\Omega}(h,m)\vec{W}(h),
\end{eqnarray}
where vector \(\vec{\Omega}(h,m)\): 
\begin{eqnarray}
\label{eq:39}
\vec{\Omega}(h,m)=(\Omega_{CC},\Omega_{CD},\Omega_{DC},\Omega_{DD}),
\end{eqnarray}
defines the probabilities of the outcomes \((CC,CD,DC,DD)\) in a competition between players \(h\) and \(m\), while \(\vec{W}(h)\) defines the corresponding payoffs (\ref{eq:1}) of \(h\). The probabilities of the outcomes \(\vec{\Omega}(h,m)\) depends on the strategies of the competitors \(h\) and \(m\) to choose their roles. 

It has become a common choice to use memory-one strategies \(\{ s_{M_{1}}\}\) as a set of possible strategies to play a two-person game\cite{Nowak1990,Hauert1997,Stewart2016}. Under the assumption of repeated rounds of the game, an iterated strategy comprises four independent probabilities: 
\begin{eqnarray}
\label{eq:15}
 s_{M_{1}}=(p_{CC},p_{CD},p_{DC},p_{DD}),
\end{eqnarray}
to cooperate if the outcome of the previous round of the game was \((CC,CD,DC,DD)\), respectively. Memory-one strategies \(\{ s_{M_{1}}\}\) form a subset of memory-N strategies \(\{ s_{M_{N}}\}\) that depend on \(N\) previous rounds of the game\cite{Hilbe2017}.  

Some memory-one strategies possess the most peculiar properties\cite{Boerlijst1997,Sigmund2010,Press2012}. For instance, an equalizer strategy \(s^{EQ}_{M_{1}}=(p_{CC},p_{DD})\) with the probabilities \(p_{CD}\) and \(p_{DC}\): 
\begin{eqnarray}
\label{eq:15986}
p_{CD}&=&\frac{p_{CC}(T-P) - (1 + p_{DD})(T-R)}{R - P},\nonumber\\
p_{DC}&=&\frac{(1 - p_{CC})(P-S) + p_{DD}(R-S)}{R - P},
\end{eqnarray}
causes the payoffs of all the opponents to be equal to each other: 
\begin{eqnarray}
\label{eq:2}
G(s_{M_{1}},s^{EQ}_{M_{1}})=const,
\end{eqnarray}
for any memory-one strategy \(s_{M_{1}}\) (the gain, however, depends on \(s^{EQ}_{M_{1}}\))\cite{Boerlijst1997,Sigmund2010}. Recently, there has been considerable interest\cite{Adami2013,Stewart2013,Hilbe2015,Stewart2016,Adami2016} in more general zero-determinant (ZD) strategies\cite{Press2012} that unilaterally set a gain (\ref{eq:44}) for the opponents.

A complex strategy may be reduced to a simple one: for instance, a memory-one strategy marginalizes any memory-N strategy. The shortest-memory player sets the outcome probabilities of a game: 
\begin{eqnarray}
\label{eq:53}
\left < \Omega_{x,y|H_{0},H_{1}}\right >_{H_{0},H_{1}}=\left < \Omega_{x,y|H_{0}}\right >_{H_{0}},
\end{eqnarray} 
where \(\Omega_{x,y}\) is a probability for specific outcome of a fight (\(x\) and \(y\) take the values \(C\) or \(D\)), while \(H_{0}\) and \(H_{1}\) represent the histories of the game available to the first and the second players correspondingly (see eq. (18) of Appendix A in \cite{Press2012}). Eq. (\ref{eq:53}) does not limited to memory-N strategies, because the variables \(H_{0}\) and \(H_{1}\) can represent any information that affects the decisions of the players.

Memory-one strategies are probably the most studied ones and are constantly in the focus of recent research\cite{Stewart2016}. Much uncertainty still exists about the impact of memory-one strategies on the evolution of animal contests. The obstacle in applying these strategies to field observations is that many mating or territory control combats lack repeated rounds, though there are some exceptions\cite{Milinski1987}. To the best of our knowledge, it is still not known whether marginalization (\ref{eq:53}) takes place in nature\cite{Lee2015}.

An example of a contest without repeated rounds in nature is the mating combat of male bowl and doily spiders\cite{Austad1983}. According to S. Austad, upon reaching maturity, specimens of male bowl and doily spiders compete with each other for access to a female's web in order to fertilize her eggs. The fight between two males proceeds as a series of aggressive grapples until one of the competitors either receives a severe injury and dies or flees to search for other mating opportunities. The winner gains access to the web. Each fight has a winner: spiders never cooperate to share the eggs of a web. The Darwinian payoffs of a spider are well-represented by the amount of inseminated eggs.

Data collected by S. Austad is sufficient to consider the mating combat of male bowl and doily spiders as a two-person game with outcome probabilities (\ref{eq:39}): 
\begin{eqnarray}
\label{eq:20}
\vec{\Omega}=(0.0,0.165,0.165,0.67),
\end{eqnarray}
and payoffs (\ref{eq:1}): 
\begin{eqnarray}
\label{eq:38}
 \vec{W}= (R=1,S=0.34,T=1.66,P=0),
\end{eqnarray}
see Appendix \ref{SEC:appspid}. The main aims of this study is understanding the link between (\ref{eq:20}) and (\ref{eq:38}) together with implications for information exchange between the spiders. 

In this article, we show that complex strategies are marginalized (\ref{eq:53}) by equalizer strategies, rather than by a general memory-one strategy. Thus, under specific circumstances, the analysis of equalizer strategies suffice to predict the outcomes of a combat. Outcome probabilities \(\vec{\Omega}\) of a combat between bowl and doily spiders (\ref{eq:20}) correspond to a unique equalizer strategy: for each payoff \(\vec{W}\) of conflict escalation type there exists a single equalizer strategy \(s^{EQnoCC}_{M_{1}}\) with no mutual cooperation in a competition between two strategies of this type (\(\Omega_{CC}=0\) in \(\vec{\Omega}(s^{EQnoCC}_{M_{1}},s^{EQnoCC}_{M_{1}})\), see (\ref{eq:39})). In addition, the strategy \(s^{EQnoCC}_{M_{1}}\) does not assess opponent's state information and, therefore, corresponds to self assessment.

The standard approach to calculate the statistics of outcomes \(\vec{\Omega}\) (\ref{eq:39}) as a function of the payoffs \(W\) requires us to choose a game, then to guess a set of the relevant strategies \(\{ h\}\), to find an evolutionarily stable strategy\cite{SMITH1973} (ESS) \(h^{ESS}\) as a function of the payoffs \(\vec{W}\), and finally to calculate the outcome probabilities \(\vec{\Omega}(h^{ESS},h^{ESS})\). The ESS strategy adopted by all members of a population hinders the survival of any mutant. Along this work, we adopt the first condition for an ESS: a strategy \(h^{ESS}\) is the ESS if its gain \(G(h^{ESS},h^{ESS})\) (\ref{eq:44}) is greater than the gain of any mutant \(m\) in a competition against \(h^{ESS}\): 
\begin{eqnarray}
\label{eq:16}
 G(h^{ESS},h^{ESS})>G(m,h^{ESS}).
\end{eqnarray}
We assume that the mutants are close to the hosts \(m\approx h\). The first condition for ESS (\ref{eq:16}) is consistent with strict Nash equilibrium. There are other definitions of ESS\cite{SMITH1973}, and there are games without ESS\cite{Sigmund1998}. 

Prediction of outcome probabilities \(\vec{\Omega}^{EQ}=\vec{\Omega}(s^{EQnoCC}_{M_{1}},s^{EQnoCC}_{M_{1}})\) with the help of equalizer strategies bypasses the calculation of the ESS of a strategy space \(\{ h\}\). As we will show, there exists a strategy space that predicts an evolutionarily stable strategy \(h^{ESS}\) with the same outcome probabilities \(\vec{\Omega}(h^{ESS},h^{ESS})=\vec{\Omega}(s^{EQnoCC}_{M_{1}},s^{EQnoCC}_{M_{1}})\) as an equalizer strategy \(s^{EQnoCC}_{M_{1}}\). We will discuss whether it is possible to refute the strategy spaces that predict other than \(\vec{\Omega}^{EQ}\) outcome probabilities. 

There is empirical evidence that some species implement self-assessment: they use only information regarding their own state, rather than the state of the competitor, to decide whether to flee or keep fighting for a territory or mating opportunity\cite{Marden1990,MestertonGibbons1996,Prenter2006,Dietemann2008,Keil2010,Percival2010,Martinez-Cotrina2014,Tsai2014}. There is evidence for the existence of a significant variety of assessment techniques in animal contests\cite{Enquist1983,Elwood2012,Rillich2007,Elias2008,Fawcett2013,Mesterton-Gibbons2014a,Guillermo-Ferreira2015,Benitez2017}, for a review, see \cite{Taylor2003,Arnott2009,Briffa2009}. The question exploring the conditions that favor the evolution of self-assessment during combat remains to be open.

Memory-one strategies make possible to estimate information transfer between the players. On the intuitive level, an iterated strategy can be reactive — it may depend only on the opponent's state \(p_{CC}=p_{DC},p_{CD}=p_{DD}\) \cite{Nowak1990,Baek2016}. In analogy with self-assessment, one can define the passive iterated strategy as: 
\begin{eqnarray}
\label{eq:22}
p_{CC}=p_{CD},p_{DC}=p_{DD},
\end{eqnarray}
that is independent of the opponent's state. 

Reduction of a strategy to an equalizer memory-one strategy by marginalization makes possible calculation of information exchange, rather than shared information, between the players. Mutual information \(I\) is a measure of available information about the state of a player under condition that the state of its opponents is known:
\begin{eqnarray}
\label{eq:19}
I=\sum_{xy=C,D}\Omega_{xy}\log\frac{\Omega_{xy}}{(\Omega_{xC}+\Omega_{xD})(\Omega_{Cy}+\Omega_{Dy})},
\end{eqnarray}
where \(\Omega_{xy}\) are the components of the outcome probabilities \(\vec{\Omega}\) (\ref{eq:39}). Mutual information is used to describe total flow of information between living organisms\cite{Tkacik2016} or an evolving organism and its environment\cite{Kussell2005,Donaldson-Matasci2010}. It includes all shared information and, therefore, in the case of two competing players, consists of assessment and another information that is available on the opponent.

Formally, transfer entropy is a measure of information transfer\cite{Schreiber2000}. Information transfer between two memory one strategies \(i\) and \(j\) (following adaptation of eq. (4) of \cite{Schreiber2000} for memory one strategies) is:
\begin{eqnarray}
\label{eq:12}
 T(i\leftarrow j)=\sum_{x'xy=C,D}\Omega_{xy}p_{x'|xy}\log\frac{p_{x'|xy}}{p_{x'|x}},
\end{eqnarray}
where \(xy\) is outcome of round \(N\) while \(x'\) is the role of the first player from round \(N+1\). Thus the probabilities \(p_{x'|xy}\) constitute a memory-one strategy (\ref{eq:15}). Conditional probability \(p_{x'|x}\) for a player to be in state \(x'\) at the round \(N+1\) if its state was \(x\) at the round \(N\) is:
\begin{eqnarray}
\label{eq:18}
p_{x'|x}=p_{x'|xC}f_{xC}+p_{x'|xD}f_{xD},f_{xy}=\frac{\Omega_{xy}}{\Omega_{xC}+\Omega_{xD}}.
\end{eqnarray} 
Transfer entropy is asymmetric: eq. (\ref{eq:12}) corresponds to assessment of \(j\)'s state by the strategy \(i\). Passive strategies (\ref{eq:22}) vanish transfer entropy and, therefore, lack assessment of the opponent's state.

For all conflict escalation contests, a unique equalizer strategy without mutual cooperation \(s^{EQnoCC}_{M_{1}}\) is a passive strategy (\ref{eq:22}). This finding, to the best of our knowledge, is the first theoretical evidence for self-assessment in iterated two person games. It corroborates the assumptions of self assessment in the previous studies, for instance war of attrition model\cite{SMITH1973,PARKER1981,Parker1974,HAMMERSTEIN1982,Bishop1978,MestertonGibbons1996,Payne1996a}.

Self-assessment is a core assumption of the war of attrition model\cite{MestertonGibbons1996}. This model is very important because its modifications constitute the main method to describe animal contests without repeated rounds\cite{HAMMERSTEIN1982,Bishop1978}. The war of attrition based models are used to fit the observed data of real fights between specimens of a species\cite{Stuart-Fox2006}, though some alternative views exist\cite{Takeuchi2016}.

According to war of attrition, two players engage in a fight. They choose times \(t_{1}\) and \(t_{2}\) to stay in the fight randomly according to probability distributions \(p_{1}(t)\) and \(p_{2}(t)\), respectively. The winner receives prize \(V\), which is independent of time. The cost of fight as a function of time is \(g(t)\), where \(g(t)\) is a monotonic function of time in the units of \(V\) (linear dependence on time is a common choice). Payoffs of the winner and the less fortunate opponent in a fight with duration \(t\) are \(V-g(t)\) and \(-g(t)\), respectively. War of attrition with the cost of fight rising linearly with time predicts the fight duration that decreases exponentially with time. War of attrition, like the snowdrift game, is a conflict escalation contest: the escalation of the fight for a long duration \(t\) results in minimal payoff for each player. 

In the case of symmetric contests, there are two possible modifications of the war of attrition model as is presented in the previous paragraph. For instance, S. Austad\cite{Austad1983} demonstrated that the data on bowl and doily spiders fit the war of attrition model with deviations. The distribution of fight duration across spiders of the same size, contrary to the predictions of the model, does not decay exponentially with time. This discrepancy indicates either a deviation from the war of attrition model's central assumption of no state assessment of the opponent\cite{Enquist1983,MestertonGibbons1996,Kim2014}  or the cost of fight rising non-linearly with time\cite{Bishop1978}. The application of iterated strategies supports the war of attrition model with self-assessment and with non-linear cost of fight as a function of time. 

\begin{figure}[htb]
\centering
\includegraphics[width=.9\linewidth]{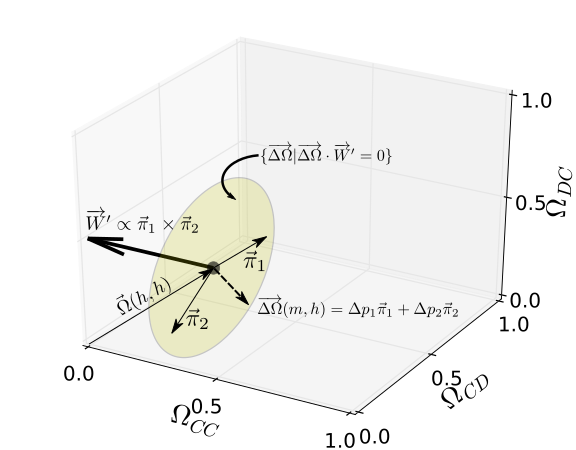}
\caption{\label{om3deqv} Competitions of an arbitrary two-parameter strategy $h(p_{1},p_{2})$ with its near-mutants $m\approx h$ may generate the same outcome probabilities $\vec{\Omega}(m,h)$ as the competitions of an equalizer strategy $s^{EQ}_{M_{1}}$ with other memory-one strategies. In linear approximation, vectors $\vec{\pi}_{1}$ and $\vec{\pi}_{2}$ define a plane $(\vec{\pi}_{1},\vec{\pi}_{2})$ in the space $\vec{\Omega}$ for outcome probabilities $\vec{\Omega}(m,h)$. An equalizer strategy with payoffs $\vec{W}'\propto \vec{\pi}_{1}\times\vec{\pi}_{2}$ and other memory-one strategies generates outcome probabilities in the form of a plane (yellow) that is parallel to $(\vec{\pi}_{1},\vec{\pi}_{2})$. These two planes coincide if $\Omega_{M_{1}}(s^{EQ}_{M_{1}}(W'),s^{EQ}_{M_{1}}(W'))=\Omega(h,h)$.}
\end{figure}

This work proceeds with the prediction of unique values for outcome probabilities (\ref{eq:39}) as a function of the evolutionary payoffs (\ref{eq:1}) and justification of self-assessment for conflict escalation contests, for instance the snowdrift game. The limitations of the findings and their implications are discussed at the end of the article. 

\section{Methods/Results}
\label{sec:org2db4c75}

Let us show that for an arbitrary two-parameter strategy \(h_{K}\)  a corresponding equalizer strategy \(s^{EQ}_{M_{1}}\)  may exist such that probabilities of the outcomes \(\Omega(h_{K},m_{k})\) are identical to probabilities of the outcomes \(\Omega(s^{EQ}_{M_{1}},s_{M_{1}})\), where  \(h_{K}\) is a function of two parameters \(h_{K}(p_{1},p_{2})\), \(m_{K}\approx h\) and \(s_{M_{1}}\)  is a memory-one strategy (not necessarily an equalizer one). 

The probability of the outcomes \(\vec{\Omega}=(\Omega_{CC},\Omega_{CD},\Omega_{DC},\Omega_{DD})\) and the corresponding payoffs \(\vec{W}=(R,S,T,P)\), despite having four components, are three-dimensional vectors: 
\begin{eqnarray}
\label{eq:14}
&&\vec{\Omega}=(\Omega_{CC},\Omega_{CD},\Omega_{DC}),\nonumber\\
&&\vec{W} = (1,S,T).
\end{eqnarray}
The vector of probabilities \(\vec{\Omega}\) has three independent parameters because \(\Omega_{CC}+\Omega_{CD}+\Omega_{DC}+\Omega_{DD}=1\). Transformation: 
\begin{eqnarray}
\label{eq:8}
\vec{W}\rightarrow \frac{\vec{W}-P}{R-P} =(1,S,T,0),
\end{eqnarray} 
changes gain \(G\rightarrow G/(R-P)-P/(R-P)\) for all competitors. Thus, transformation (\ref{eq:8}) does not affect the relative gains (\ref{eq:12}) and, therefore, does not affect the ESS condition (\ref{eq:16}). Using (\ref{eq:8}), the average gain of player \(p\) against a player \(q\) is: 
\begin{eqnarray}
\label{eq:30}
&&G(p,q)=\\\nonumber
&&1\times\Omega_{CC}+S\times\Omega_{CD}+T\times\Omega_{DC}+0\times\Omega_{DD}=\vec{W}\vec{\Omega},
\end{eqnarray}
where \(\vec{\Omega}\) and \(\vec{W}\) are defined by (\ref{eq:14}). The equalizer strategies (\ref{eq:15986}) remain unaffected by the transformation (\ref{eq:8}). 

The outcome probabilities \(\vec{\Omega}(s_{M_{1}},s^{EQ}_{M_{1}})\) of the competitions between an equalizer strategy \(s^{EQ}_{M_{1}}\) and other memory-one strategies \(s_{M_{1}}\) form a two-dimensional plane in the three-dimensional space \(\vec{\Omega}\) (\ref{eq:14}), see Figure \ref{om3deqv}. All strategies \(s_{M_{1}}\) possess the same payoff against an equalizer strategy \(s^{EQ}_{M_{1}}\). Following (\ref{eq:2}) and (\ref{eq:44}), \(G(s_{M_{1}},s^{EQ}_{M_{1}})=\vec{\Omega}(s_{M_{1}},s^{EQ}_{M_{1}})\cdot \vec{W}=\text{const}\). Thus, \(\vec{\Omega}(s_{M_{1}},s^{EQ}_{M_{1}})\) forms a plane that is perpendicular to \(\vec{W}\) and passes the point \(\vec{\Omega}(s^{EQ}_{M_{1}},s^{EQ}_{M_{1}})\): \(\vec{\Omega}(s_{M_{1}},s^{EQ}_{M_{1}})=\vec{\Omega}(s^{EQ}_{M_{1}},s^{EQ}_{M_{1}})+\Delta\vec{\Omega}\) and \(\Delta\vec{\Omega}\cdot \vec{W}=0\). 

For an arbitrary strategy \(h_{K}=(p^{h}_{1},p^{h}_{2})\), the linear expansion of outcome probabilities \(\vec{\Omega}_{K}(p^{m}_{1},p^{m}_{2},p^{h}_{1},p^{h}_{2})\) between \(h_{K}\) and its mutants \(m_{K}=(p^{m}_{1},p^{m}_{2})\) defines a plane in the \(\vec{\Omega}\) space: 
\begin{eqnarray}
\label{eq:42766}
&&\vec{\Delta\Omega}(m,h)=\\\nonumber &&\vec{\Omega}_{K}(p^{m}_{1},p^{m}_{2},p^{h}_{1},p^{h}_{2})-\vec{\Omega}_{K}(p^{h}_{1},p^{h}_{2},p^{h}_{1},p^{h}_{2})=\\\nonumber
&&\vec{\pi}_{1}\Delta p_{1}+\vec{\pi}_{2}\Delta p_{2},
\end{eqnarray}
where \(\vec{\pi}_{1}\) and \(\vec{\pi}_{2}\) are the vectors in the \(\vec{\Omega}\) space: 
\begin{eqnarray}
\label{eq:47}
&&\vec{\pi}_{1}=\left .\frac{\partial\vec{\Omega}(p_{1}^{m},p_{2}^{m},p_{1},p_{2})}{\partial p_{1}^{m}}\right |_{\substack{p^{m}_{1}=p^{h}_{1} \\ p^{m}_{2}=p^{h}_{2} }},\nonumber\\
&&\vec{\pi}_{2}=\left .\frac{\partial\vec{\Omega}(p_{1}^{m},p_{2}^{m},p_{1},p_{2})}{\partial p_{2}^{m}}\right |_{\substack{p^{m}_{1}=p^{h}_{1} \\ p^{m}_{2}=p^{h}_{2} }},\end{eqnarray}
Thus, \(\vec{\Delta\Omega}(m,h)\) forms a plane perpendicular to vector \(\vec{W}'\): 
\begin{eqnarray}
\label{eq:56}
\{\vec{\Delta\Omega}|\vec{\Delta\Omega}\cdot \vec{W}'=0\},
\end{eqnarray}
where: 
\begin{eqnarray}
\label{eq:60}
\vec{W}'\propto\vec{\pi}_{1}\times\vec{\pi}_{2},
\end{eqnarray}
see Figure \ref{om3deqv}. 

\begin{figure}[htb]
\centering
\includegraphics[width=.9\linewidth]{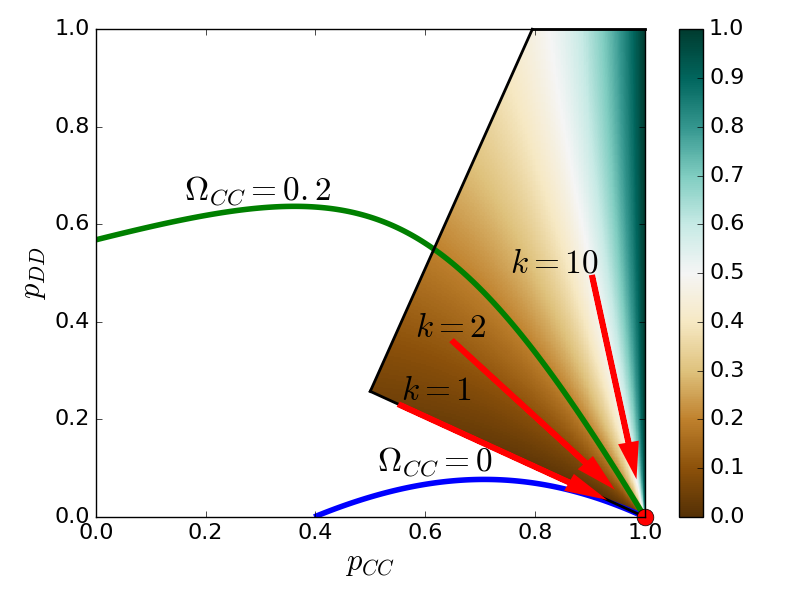}
\caption{\label{Occeqv} Probability of mutual cooperation $\Omega_{CC}$ in a competition between two identical equalizer strategies $(p_{CC},p_{DD})$ for a snowdrift game. Color map of $\Omega_{CC}$ values indicate allowed equalizer strategies. There exists a single competition without mutual cooperation and valid strategies, see the intersection between the contour line for competitions without mutual cooperation $\Omega_{CC}=0$ and the region for valid strategies (line $\Omega_{CC}=0.2$ is brought to show that only $\Omega_{CC}=0$ possesses a unique allowed strategy). No mutual cooperation occurs at the strategy $(p_{CC},p_{DD})=(1,0))$. Probability $\Omega_{CC}$ has a singularity at this strategy and the limits of outcome probabilities $\vec{\Omega}$ depend on the direction of approach to this point. The arrows indicate different directions $p_{DD}=k (-S + p_{CC} S)/(-1 + S)$. No mutual cooperation corresponds to the direction along the left boundary of the allowed strategies region $k=1$. The limit $\vec{\Omega}$ without mutual cooperation fits the observations of the mating combats of male bowl and doily spiders. The corresponding memory-one strategy is $(p_{CC},p_{CD},p_{DC},p_{DD})=(1,1,0,0)$. This memory-one strategy is passive $(p_{CC}=p_{CD},p_{DC}=p_{DD})$. Passive strategies ignore information about the opponent's state.}
\end{figure}

The planes formed by mutants of \(h_{K}\) and the equalizer strategy \(s^{EQ}_{M_{1}}\) coincide if: 
\begin{eqnarray}
\label{eq:28}
\Omega_{M_{1}}(s^{EQ}_{M_{1}}(W'),s^{EQ}_{M_{1}}(W'))=\Omega_{K}(h_{K},h_{K}),\\\nonumber
\end{eqnarray}
where \(W'\) is defined by (\ref{eq:60}) and (\ref{eq:47}). Eq. (\ref{eq:28}) suggests that only memory-one strategies depend on the payoffs (\ref{eq:15986}). 

Equalizer strategy \(s^{EQ}_{M_{1}}(W')\) marginalizes (\ref{eq:53}) strategy \(h_{K}\) in all competitions between \(h_{K}\) and its near mutants \(m\approx h_{K}\) if both eqs. (\ref{eq:60}) and (\ref{eq:28}) hold. If only (\ref{eq:28}) holds then \(s^{EQ}_{M_{1}}(W')\) marginalizes strategy \(h_{K}\) only in the contest against itself. 

The payoffs \(\vec{W}'\) in (\ref{eq:60}) and (\ref{eq:28}) may and may not coincide with the payoffs \(\vec{W}\) (\ref{eq:8}) that define the evolutionary dynamics for the strategies \(h_{K}\). Thus, there are three possibilities for a solution \(s^{EQ}_{M_{1}}(h_{K})\) to eq. (\ref{eq:28}). First, there exists a memory-one strategy \(s^{EQ}_{M_{1}}\) that fits eqs. (\ref{eq:60}) and (\ref{eq:28}) and: 
\begin{eqnarray}
\label{eq:46}
\vec{W}'\propto W.
\end{eqnarray}
Proportionality of the payoffs \(\vec{W}'\) and \(\vec{W}\) requires: 
\begin{eqnarray}
\label{eq:64}
&&\left. \nabla_{m_{K}}G(m_{K},h_{K})\right |_{m_{K}=h_{K}}=\nonumber\\
&&\vec{W}\cdot\left. \nabla_{m_{K}}\Omega(m_{K},h_{K})\right |_{m_{K}=h_{K}}=0,
\end{eqnarray}
Condition (\ref{eq:64}) holds for some ESS \(h^{ESS}_{K}\) (\ref{eq:16}): the first presentation of equalizer strategies indicated connection of these strategies to the notion of an ESS\cite{Boerlijst1997,Hilbe2015}. Second, eqs. (\ref{eq:46}) and (\ref{eq:28}) have a solution and: 
\begin{eqnarray}
\label{eq:51}
 \vec{W}'\neq \vec{W}.  
\end{eqnarray}
In this case, \(h_{K}\) possesses a corresponding  equalizer strategy \(s^{EQ}_{M_{1}}(h_{K})\) with payoff \(W'\) that is incapable of being justified or explained, because \(\vec{W}'\) is different from the observed payoffs. Nevertheless, marginalization addresses only outcome probabilities \(\vec{\Omega}\) and therefore payoffs \(\vec{W}'\) can be considered as free parameters of the strategy. This case will be important for the discussion of self-assessment. Third, eqs. (\ref{eq:51}) and (\ref{eq:28}) have no solution for any \(\vec{W}'\). In this case, competitions of the strategy \(h_{K}\) with its mutants cannot be marginalized by a single equalizer strategy.  

\begin{figure}[htb]
\centering
\begin{tabular}{c}
\multicolumn{1}{l}{{\bf\sf A}} \\ 
\includegraphics[width=.9\linewidth]{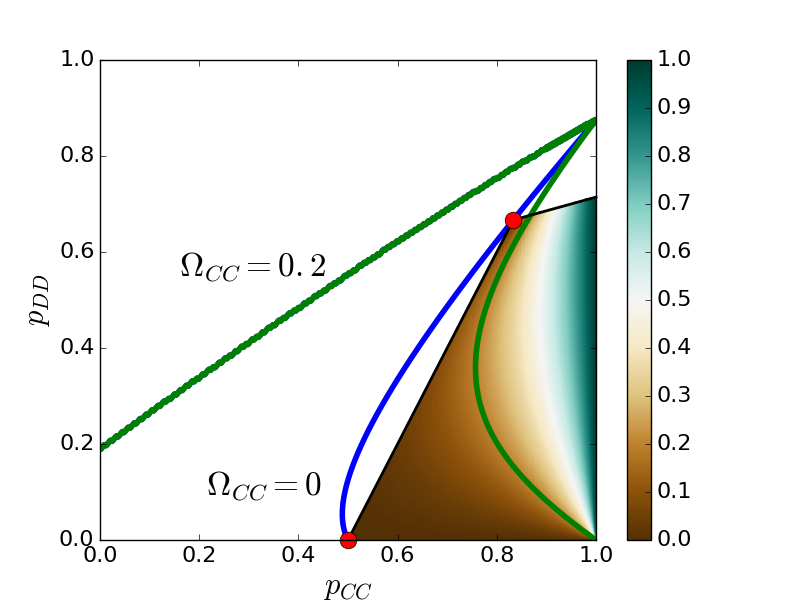} \\ 
\multicolumn{1}{l}{{\bf\sf B}} \\ 
\includegraphics[width=.9\linewidth]{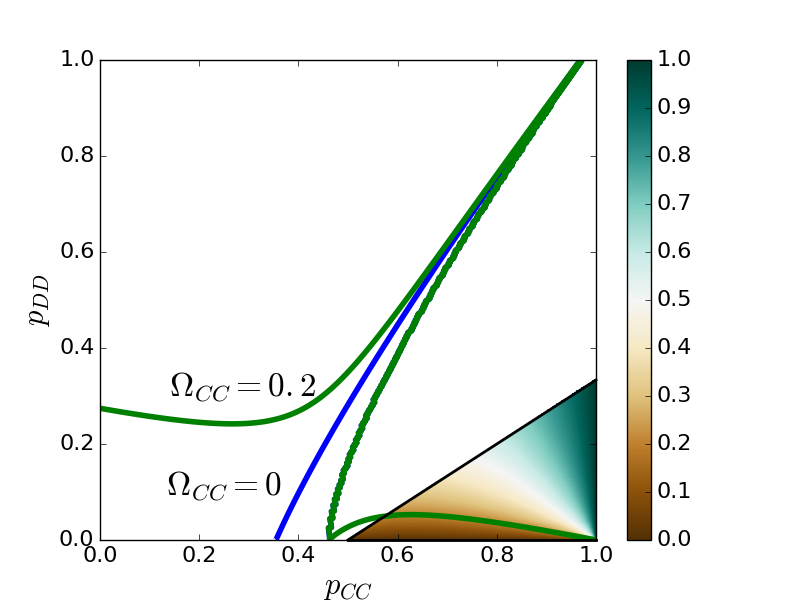} 
\end{tabular}
\caption{\label{Occeqv1} Prisoner's dilemma: probability of mutual cooperation $\Omega_{CC}$ as a function of equalizer strategy $(p_{CC},p_{DD})$. Color map of $\Omega_{CC}$ values indicates allowed equalizer strategies. A) If $S>-T$, there are two valid strategies without mutual cooperation $\Omega_{CC}=0$. The corresponding probabilities of the outcomes $\vec{\Omega}=(0,0,0,1)$ and $\vec{\Omega}=(0,0.5,0.5,0)$ are independent of the payoffs. B) If $S<-T$, there are no valid strategies with $\Omega_{CC}=0$.}
\end{figure}

For a conflict escalation \((0\leq S\leq 1,S\leq T)\): First, if eq. (\ref{eq:28}) has a solution for an arbitrary memory-one strategy \(s_{M_{1}}\), then this strategy is an equalizer one, see Appendix \ref{SEC:appit} and Appendix \ref{SEC:appuniq}. Thus, equalizer strategies are unique in their ability to fit interactions of an arbitrary strategy with its mutants. Second, there exists a single equalizer strategy \(s^{EQnoCC}_{M_{1}}\) which predicts no mutual cooperation \(\Omega_{CC}=0\) in the probabilities of the outcomes \(\vec{\Omega}(s^{EQnoCC}_{M_{1}},s^{EQnoCC}_{M_{1}})\): 
\begin{eqnarray}
\label{eq:65755}
\vec{\Omega}(s^{EQnoCC}_{M_{1}},s^{EQnoCC}_{M_{1}})=(0,\frac{S}{S+T},\frac{S}{S+T},\frac{T-S}{S+T}),
\end{eqnarray}
see Appendix \ref{SEC:appit} for a detailed derivation. The graphical presentation in Figure \ref{Occeqv} shows that (\ref{eq:65755}) holds in the limit: 
\begin{eqnarray}
\label{eq:17}
 s^{EQnoCC}_{M_{1}}=(p_{CC},p_{DD})\rightarrow (1,0),
\end{eqnarray}
with the corresponding memory-one iterated strategy \(s_{M_{1}}\) 
\begin{eqnarray}
\label{eq:32}
 s_{M_{1}}\rightarrow (1,1,0,0),
\end{eqnarray}
see (\ref{eq:15986}). The outcome probabilities have a singularity in the limit (\ref{eq:17}). The limit corresponds to different values of \(\Omega\). Nevertheless, only single \(\Omega\) lacks mutual cooperation. Third, the equalizer strategy \(s^{EQnoCC}_{M_{1}}\) (\ref{eq:17}) is passive (\ref{eq:22}). Thus it does not assess information of its mutants' states.

Marginalization (\ref{eq:28}) may be applied only to a two parameter strategy. Two parameter strategies \(h_{K}(p_{1},p_{2})\) suffice to describe outcome probabilities \(\vec{\Omega}\). In a competition of two identical strategies \(h_{K}(p_{1},p_{2})\) there are two independent outcome probabilities \(\Omega_{CC}\) and \(\Omega_{CD}\) (\(\Omega_{CD}=\Omega_{DC}\) in (\ref{eq:14})) and two independent strategy parameters \((p_{1},p_{2})\). Four independent strategy parameters in a competition of two different strategies \(h_{K}(p_{1},p_{2})\) and \(h'_{K}(p'_{1},p'_{2})\) even overfit three independent outcome probabilities (\ref{eq:14}). 

The prediction of outcome probabilities (\ref{eq:65755}) is limited to a conflict escalation, e.g. the snowdrift game. Equalizer strategies in the case of prisoner’s dilemma payoffs do not possess outcome probabilities (\ref{eq:65755}), see Figure \ref{Occeqv1} and Appendix \ref{SEC:appit}. 

Probabilities of the outcomes (\ref{eq:65755}) fit the observations of bowl and doily spiders surprisingly well, see Appendix \ref{SEC:appspid}. By substituting weights (\ref{eq:38}) for (\ref{eq:65755}) one gets: 
\begin{eqnarray}
\label{eq:3}
\vec{\Omega}=(0,0.17,0.17,0.66),
\end{eqnarray} 
which is about \(1\%\) close to the values (\ref{eq:20}) (unfortunately, no error bars are available). In addition, self-assessment was assumed in a treatment of bowl and doily spiders using the war of attrition model. 

\section{Discussion}
\label{sec:orgad68546}

The universality of outcome probabilities (\ref{eq:65755}) and predictions of self-assessment depend on whether any relevant ESS strategy is, necessarily, marginalized (\ref{eq:53}) by an equalizer strategy. This question is part of a more general discussion regarding the relevant strategies to describe behavior of species during a competition. For instance, can we refute a strategy that predicts different from (\ref{eq:65755}) outcome probabilities?

\begin{figure}[htb]
\centering
\includegraphics[width=.9\linewidth]{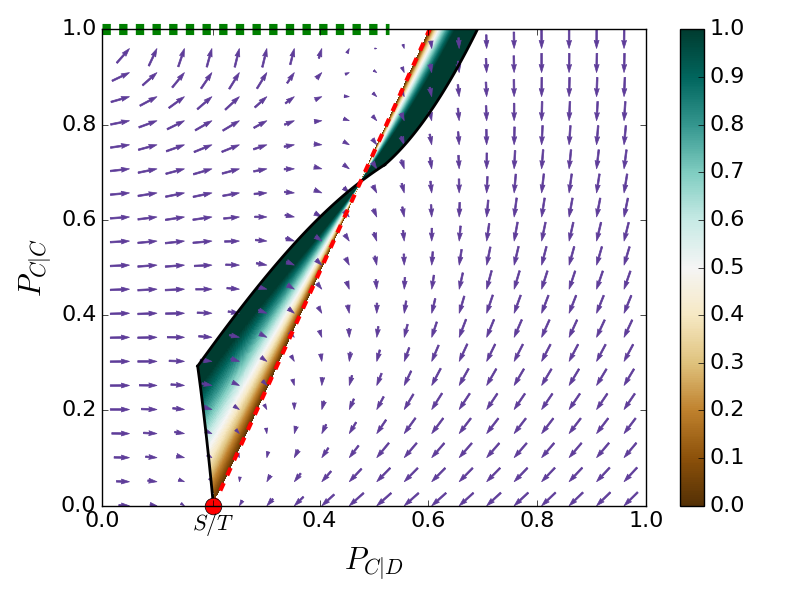}
\caption{\label{RepDynSD}
Snowdrift game: evolutionary stability of the predicted strategy $(S/T,0)$ in the $(P_{C|D},P_{C|C})$ strategy space. Adaptive dynamics fluxes (blue arrows) converge to the predicted state at $(S/T,0)$ (red point) from  the majority of possible initial coordinates. Continuous region (green dashed line) of convergence is a property of $(P_{C|D},P_{C|C})$ strategy space without a matching equalizer strategy. The strategy $(S/T,0)$ belongs to the subspace of iterated equalizer zero-determinant strategies (colored region), compare with Figure \ref{Occeqv}. The colors indicate proximity to the $(1,1,0,0)$ memory-one strategy, where brown color corresponds to close strategies and blue color corresponds to distant strategies. Red dashed line corresponds to the equalizer strategies in the limit $(1,1,0,0)$, see red circle in Figure \ref{Occeqv}. Only the strategy $(S/T,0)$ lacks mutual cooperation $P_{C|C}=0$.}
\end{figure}

To support outcome probabilities (\ref{eq:65755}), let us the generalize the strategies by elimination of \(H_{0}\) in (\ref{eq:53}). Further generalization is required because \(H_{0}\) is game-dependent. The elimination of \(H_{0}\) may lead to the description of the strategies and the outcome probabilities \(\vec{\Omega}\) that are game-independent and, therefore, universal. 

To eliminate \(H_{0}\) in (\ref{eq:53}), let us express the joint outcome probabilities \(\vec{\Omega}\) (\ref{eq:39}) as a function of conditional probabilities \(P_{C|D}\) and \(P_{C|C}\). The standard notation of conditional probability \(P_{A|B}\) for \(A\) under the condition of \(B\) is used. For instance, \(P_{C|D}\) is a probability to cooperate against a defector. A single player's strategy, then, is: 
\begin{eqnarray}
\label{eq:21}
 h_{K}=(P_{C|D},P_{C|C}).
\end{eqnarray}
The conditional probabilities to be \(D\) vs. \(D\) and to be \(C\) vs \(D\) are \(P_{D|D}=1-P_{C|D}\) and \(P_{D|C}=1-P_{C|C}\), respectively. Strategies with conditional probabilities were used in sequential assessment games (see page 403 of \cite{Enquist1983}) and in herding models\cite{Kirman1993}.

In a competition of two players \(i\) and \(j\), the joint probabilities of the outcomes \(\vec{\Omega}\) (\ref{eq:39}) and the conditional probabilities \((P^{i}_{C|D},P^{i}_{C|C})\) and \((P^{j}_{C|D},P^{j}_{C|C})\) fit the following relations: 
\begin{eqnarray}
\label{eq:59}
P^{i}_{C|C}=\frac{\Omega_{CC}}{\Omega_{DC}+\Omega_{CC}},P^{j}_{C|C}=\frac{\Omega_{CC}}{\Omega_{CD}+\Omega_{CC}},\\\nonumber
P^{i}_{C|D}=\frac{\Omega_{CD}}{\Omega_{CD}+\Omega_{DD}},P^{j}_{C|D}=\frac{\Omega_{DC}}{\Omega_{DC}+\Omega_{DD}},
\end{eqnarray}
This system always has a solution for conditional probabilities as a function of outcome probabilities. Outcome probabilities can be expressed as a function of the conditional probabilities only under specific circumstances, because of constraint \(\Omega_{CC}+\Omega_{CD}+\Omega_{DC}+\Omega_{DD}=0\) and due to omitted causality (who is the first and who is the second to respond during the competition).

The ESS strategy (\ref{eq:21}) under the condition of no mutual cooperation (\(P_{C|C}=0\) for all players) predicts the same outcome probabilities as can be predicted with the help of equalizer strategies (\ref{eq:65755}). The solution of (\ref{eq:59}) if \(P^{i}_{C|C}=P^{j}_{CC}=0\) is: 
\begin{eqnarray}
\label{eq:61}
&&\Omega_{CC}=0,\\\nonumber
&&\Omega_{CD}=\frac{P^{i}_{C|D}-P^{i}_{C|D}P^{j}_{C|D}}{1-P^{i}_{C|D}P^{j}_{C|D}},\\\nonumber
&&\Omega_{DC}=\frac{P^{j}_{C|D}-P^{j}_{C|D}P^{i}_{C|D}}{1-P^{i}_{C|D}P^{j}_{C|D}}.\\\nonumber
\end{eqnarray}
These equations are valid unless there are constraints on the possible values of conditional probability \(P_{C|D}\).

The corresponding ESS strategy (\ref{eq:16}): 
\begin{eqnarray}
\label{eq:29}
h^{ESS}_{K}=(\frac{S}{T},0),
\end{eqnarray}
follows the substitution of (\ref{eq:61}) in the gain (\ref{eq:30}) and solving (\ref{eq:16}) for \(h^{ESS}\). Finally, the substitution of (\ref{eq:29}) in (\ref{eq:61}) results in the outcome probabilities (\ref{eq:65755}). Thus the validity of (\ref{eq:65755}) breaks down only if eqs. (\ref{eq:61}) no longer hold.

Outcome probabilities (\ref{eq:65755}) are predicted with the help of equalizer strategies and with the help of strategies (\ref{eq:29}). Both derivations assume the lack of mutual cooperation \(\Omega_{CC}=0\). Releasing this assumption may affect the stability of the predicted strategy.

To show that general evolutionary stability of (\ref{eq:29}) and (\ref{eq:65755}) is possible, let us present an example of a competition that possesses an evolutionarily stable strategy (\ref{eq:29}) even if mutual cooperation \(\Omega_{CC}\neq 0\) takes place. We leave out of the scope of this work all predictions of this specific example but the stability of (\ref{eq:29}). 

The competition is an iterated two-person game. During a single round of the game, the players are assigned to be the first (non-rational behavior) and the second (rational behavior). The first player chooses role \(D\) with probability \(P_{D}\) or role \(C\) with probability \(1-P_{D}\), irrespective of the opponent's state. The second player chooses role \(C\) with the conditional probability \(P_{C|C}\) if the state of the opponent is \(C\), and chooses role \(C\) with the conditional probability \(P_{C|D}\) if the opponent is in state \(D\). Then, the players receive their payoffs (\ref{eq:1}). The total gain of a player is defined over multiple rounds of the game, where each player has equal probabilities to be either the first or the second. 

\begin{figure}[htb]
\centering
\includegraphics[width=.9\linewidth]{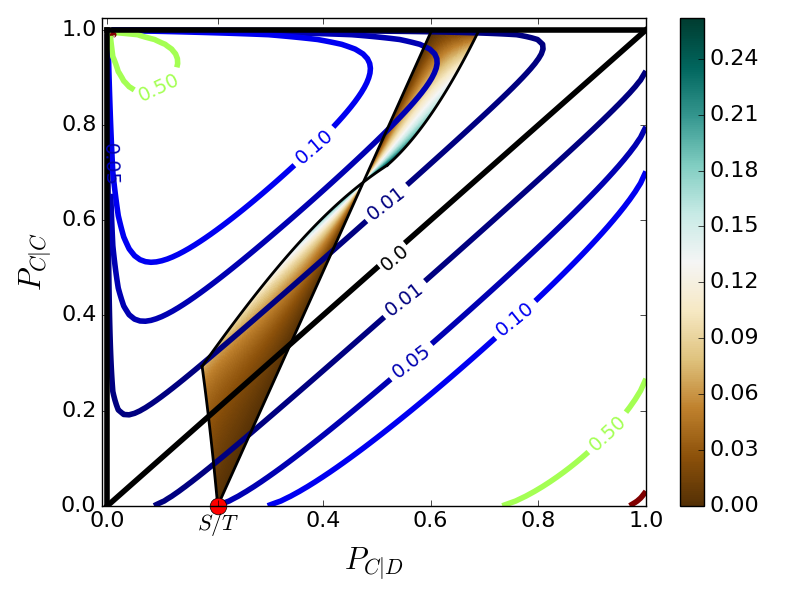}
\caption{\label{MTinfo} Transfer entropy as a function of equalizer strategies and mutual information as a function of $(P_{C|D},P_{C|C})$ strategies. Mutual information (contour lines) includes all possible channels of information exchange and its processing: assessment, previous knowledge and external signals. It depends only on the outcome probabilities of the competition. Transfer entropy (color map) is a measure of information flow between the players and, therefore, is a better estimate for assessment of the opponent's state. It can be calculated only for memory one strategies, e.g. equalizer strategies. Transfer entropy vanishes at the strategy $(S/T,0)$ because corresponding equalizer strategy is passive. Mutual information at the strategy $(S/T,0)$ keeps finite value.}
\end{figure}

To calculate the outcome probabilities (\ref{eq:14}) as a function of strategies (\ref{eq:21}), we assume that probability \(P_{D}\) of a player (\ref{eq:21}) to be in state \(D\) is the same during both rational and non-rational behavior. This assumption that the probability distribution of a variable without any additional external information (non-rational player does not know the state of the opponent) should be the same as its average behavior. 

In the case of the snowdrift game \((0<S<1,T>1)\), Figure \ref{RepDynSD} demonstrates the relation between the \(h_{K}\) strategy space (\ref{eq:21}) and equalizer memory-one strategies \(s_{M_{1}}\) (\ref{eq:15}). The colored region covers the \(K\) strategies that possess a corresponding equalizer strategy with the same probabilities of outcomes (\ref{eq:28}) for \(\vec{W}'=\vec{W}\), see Appendix \ref{SEC:appnonit} eq. (\ref{eq:1735}). Color indicates the distance of the corresponding memory-one strategy from \((1,1,0,0)\) , which is a unique passive strategy in this space. This strategy corresponds to the point (\ref{eq:29}) (red point) with outcome statistics (\ref{eq:65755}). Indeed, it is a single equalizer strategy without mutual cooperation \(P_{C|C}=0\). 

The strategy (\ref{eq:29}) is ESS for the entire range of snowdrift game payoffs \((0<S<1,T>1)\). The vector flow in Figure \ref{Occeqv} is adaptive dynamics\cite{Sigmund1998} (\ref{eq:27}) in the \(K\) space that converges to this strategy (there is an additional region of convergence (red line) that is specific to the \(K\) space and is, therefore, out of the scope of this work). The ESS (\ref{eq:29}) is of the first type (\ref{eq:16}) which is, in general, consistent with the strict Nash equilibrium. It is important to note that the evolutionary stability of (\ref{eq:29}) depends on the assumption of small mutant steps. Condition (\ref{eq:64}) does not hold for \(T>1\), but holds for \(T<1\) when mutual cooperation is better than any type of defection. Surprisingly, it makes (\ref{eq:29}) unstable: in this case, the flux along the \(P_{C|C}\) axis changes sign at (\ref{eq:29}) and, therefore, a small fluctuation of strategy takes the population out of the \(P_{C|C}=0\) boundary. 

The ESS (\ref{eq:29}) in contests with its near mutants, is marginalized by equalizer strategy (\ref{eq:28}) of passive type, see (\ref{eq:22}) and (\ref{eq:32}). The corresponding payoffs (\ref{eq:60}) are: 
\begin{eqnarray}
\label{eq:4}
\vec{W}' = (1,\frac{S}{S+T},\frac{T}{S+T}),
\end{eqnarray}
see Appendix \ref{SEC:appnonit} for the details. The payoffs (\ref{eq:4}) are different from the payoffs \(W\) (\ref{eq:8}) because condition (\ref{eq:64}) does not hold. This result is unique - only an equalizer strategy can marginalize interactions of a strategy with its mutants and there is single equalizer strategy with payoffs (\ref{eq:4}) that predicts outcome probabilities without mutual cooperation \(\Omega_{CC}=0\).

Strategy (\ref{eq:29}) possesses finite mutual information (\ref{eq:19}) but zero information transfer (\ref{eq:12}) calculated with corresponding equalizer strategy (\ref{eq:17}), see Figure \ref{MTinfo}. Mutual information may be finite even in the case of self assessment because self assessment allows some shared information. For instance, even during war of attrition with self assessment a player stops fighting when the opponents surrenders. 

Let us analyze Bowl and Doily spiders with the help of previous results. Strategy space (\ref{eq:21}) possesses ESS (\ref{eq:29}) and corresponding outcome probabilities (\ref{eq:65755}) in accord with observations (\ref{eq:20}) and (\ref{eq:38}). The ESS strategy (\ref{eq:29}) assesses the state of the opponent - a player with this strategy possesses probability \(S/T\) to be \(C\) if its opponent is \(D\). It is impossible without any information regarding the opponent's state. Nevertheless, the strategy (\ref{eq:29}) is marginalized by the equalizer strategy (\ref{eq:32}) with payoffs (\ref{eq:4}), see Figures \ref{om3deqv} and \ref{Occeqv}. At the limit (\ref{eq:32}) information exchange (\ref{eq:12}) vanishes. Thus we conclude that contest of bowl and doily spiders includes only self assessment.

\begin{figure}[htb]
\centering
\includegraphics[width=.9\linewidth]{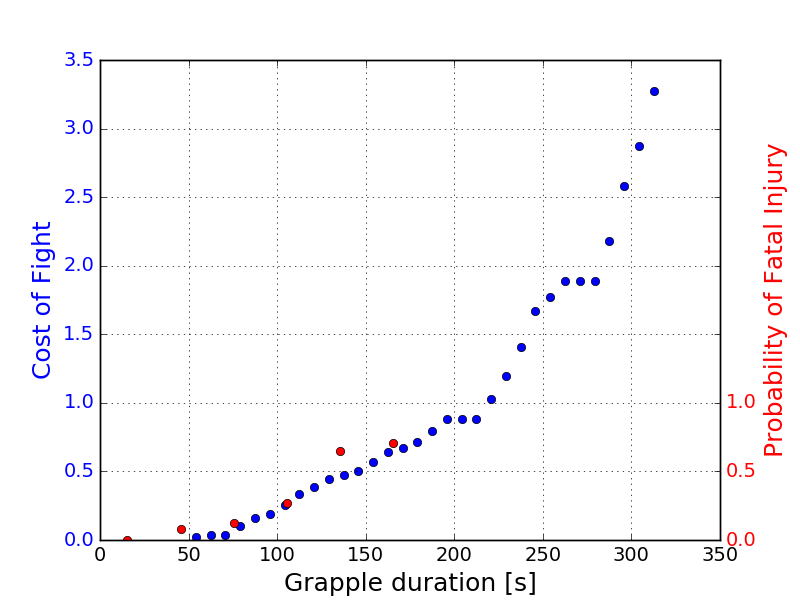}
\caption{\label{figWA}
Normalized cost of fight as a function of time (blue) multiplied by factor 2 ($2g(t)/V$) and the probability of fatal injury (red) for the combat of male bowl and doily spiders. The data of probability to suffer a fatal injury is taken according to Figure 4 from reference \cite{Austad1983} for combatants of similar size. The cost of fight is calculated from the distribution of fight duration (Figure 6 of the same reference) with the help of eq. (\ref{eq:32736}). The probability approaches its maximum value $1$ after 200 seconds. Until this time, the probability of fatal injury (red markers) and twice the cost of the fight (blue markers) fit each other. Factor two corresponds to the probability that a fight results in an injury of one of the competitors is twice the probability of a competitor getting an injury.}
\end{figure}

To fit the observed distributions of fight durations between male bowl and doily spiders of the same size (Figure 6 of \cite{Austad1983}) this work supports war of attrition with self assessment and the non-linear cost of fight. In the case of bowl and doily spiders, the predicted non-linear cost of a fight is correlated with the probability of injury during the fight, see Figure \ref{figWA} and Appendix \ref{SEC:appwaratt}. This reasonable results provides additional indirect support for the approach of this work.

Data of bowl and doily spiders fit outcome probabilities (\ref{eq:65755}) that are predicted by equalizer strategies. As a consequence it supports validity of eqs. (\ref{eq:61}) and that marginalization of the complex strategies by equalizer ones takes place in nature. Strategy space(\ref{eq:21}) provides only an example of a ESS and its marginalization by an equalizer strategy.

Marginalization of the strategies (\ref{eq:53}) can be supported only by experimental results because it does not change the outcome probabilities (\ref{eq:53}) and, therefore, does not affect the gain (\ref{eq:44}) and corresponding evolutionary stability. Further attempts to support or refute prediction of outcome probabilities (\ref{eq:65755}) and self assessment for conflict escalation, require more data of animal contests in the same vein as in S. Austad’s study on bowl and doily spiders: contest of two speciments with know outcome probabilities and evolutionary payoffs. This data is hard to find, though conflict escalation, snowdrift game\cite{Doebeli2005,Gore2009,Perc2010} or war of attrition, are quite common in nature.

To conclude, the main result of this work is the prediction of general probabilities for outcomes of a conflict escalation contest. The impetus for this study stems from a finding that there is a single equalizer strategy for a snowdrift two-person competition without mutual cooperation. This single strategy predicts observations of combats between male bowl and doily spiders surprisingly well. In addition, a player using this equalizer strategy does not assess the opponent's state (self-assessment) during competitions. This work extends the ability of iterated strategies to describe non-iterated animal contests, support that evolution reduces complex strategies to a simpler ones and contributes to the recent line of research in making evolutionary predictions by looking for general unequaled properties of the strategies instead of a game-specific analysis of evolutionary stability. 

\section{Bibliography}
\label{sec:orgb9bd8ef}


\begin{thebibliography}{60}
\expandafter\ifx\csname natexlab\endcsname\relax\def\natexlab#1{#1}\fi
\expandafter\ifx\csname bibnamefont\endcsname\relax
  \def\bibnamefont#1{#1}\fi
\expandafter\ifx\csname bibfnamefont\endcsname\relax
  \def\bibfnamefont#1{#1}\fi
\expandafter\ifx\csname citenamefont\endcsname\relax
  \def\citenamefont#1{#1}\fi
\expandafter\ifx\csname url\endcsname\relax
  \def\url#1{\texttt{#1}}\fi
\expandafter\ifx\csname urlprefix\endcsname\relax\def\urlprefix{URL }\fi
\providecommand{\bibinfo}[2]{#2}
\providecommand{\eprint}[2][]{\url{#2}}

\bibitem[{\citenamefont{Hardy and Briffa}(2013)}]{Hardy2013}
\bibinfo{author}{\bibfnamefont{I.~C.} \bibnamefont{Hardy}} \bibnamefont{and}
  \bibinfo{author}{\bibfnamefont{M.}~\bibnamefont{Briffa}},
  \emph{\bibinfo{title}{Animal contests}} (\bibinfo{publisher}{Cambridge
  University Press}, \bibinfo{year}{2013}).

\bibitem[{\citenamefont{Milinski}(2014)}]{Milinski2014}
\bibinfo{author}{\bibfnamefont{M.}~\bibnamefont{Milinski}},
  \bibinfo{journal}{Behav. Ecol.} \textbf{\bibinfo{volume}{25}},
  \bibinfo{pages}{680} (\bibinfo{year}{2014}).

\bibitem[{\citenamefont{Taylor and Elwood}(2003)}]{Taylor2003}
\bibinfo{author}{\bibfnamefont{P.~W.} \bibnamefont{Taylor}} \bibnamefont{and}
  \bibinfo{author}{\bibfnamefont{R.~W.} \bibnamefont{Elwood}},
  \bibinfo{journal}{Anim. Behav.} \textbf{\bibinfo{volume}{65}},
  \bibinfo{pages}{1195} (\bibinfo{year}{2003}).

\bibitem[{\citenamefont{Arnott and Elwood}(2009)}]{Arnott2009}
\bibinfo{author}{\bibfnamefont{G.}~\bibnamefont{Arnott}} \bibnamefont{and}
  \bibinfo{author}{\bibfnamefont{R.~W.} \bibnamefont{Elwood}},
  \bibinfo{journal}{Anim. Behav.} \textbf{\bibinfo{volume}{77}},
  \bibinfo{pages}{991} (\bibinfo{year}{2009}).

\bibitem[{\citenamefont{Briffa and Elwood}(2009)}]{Briffa2009}
\bibinfo{author}{\bibfnamefont{M.}~\bibnamefont{Briffa}} \bibnamefont{and}
  \bibinfo{author}{\bibfnamefont{R.~W.} \bibnamefont{Elwood}},
  \bibinfo{journal}{Anim. Behav.} \textbf{\bibinfo{volume}{77}},
  \bibinfo{pages}{759} (\bibinfo{year}{2009}).

\bibitem[{\citenamefont{Fawcett and Mowles}(2013)}]{Fawcett2013}
\bibinfo{author}{\bibfnamefont{T.~W.} \bibnamefont{Fawcett}} \bibnamefont{and}
  \bibinfo{author}{\bibfnamefont{S.~L.} \bibnamefont{Mowles}},
  \bibinfo{journal}{Anim. Behav.} \textbf{\bibinfo{volume}{86}},
  \bibinfo{pages}{E1} (\bibinfo{year}{2013}).

\bibitem[{\citenamefont{Mesterton-Gibbons
  et~al.}(2014)\citenamefont{Mesterton-Gibbons, Karabiyik, and
  Sherratt}}]{Mesterton-Gibbons2014}
\bibinfo{author}{\bibfnamefont{M.}~\bibnamefont{Mesterton-Gibbons}},
  \bibinfo{author}{\bibfnamefont{T.}~\bibnamefont{Karabiyik}},
  \bibnamefont{and} \bibinfo{author}{\bibfnamefont{T.~N.}
  \bibnamefont{Sherratt}}, \bibinfo{journal}{Dyn. Games Appl.}
  \textbf{\bibinfo{volume}{4}}, \bibinfo{pages}{407} (\bibinfo{year}{2014}).

\bibitem[{\citenamefont{Smith~Maynard}(1982)}]{Smith1982}
\bibinfo{author}{\bibfnamefont{J.}~\bibnamefont{Smith~Maynard}},
  \emph{\bibinfo{title}{Evolution and the Theory of Games}}
  (\bibinfo{publisher}{Cambridge university press}, \bibinfo{year}{1982}).

\bibitem[{\citenamefont{Axelrod and Hamilton}(1981)}]{AXELROD1981}
\bibinfo{author}{\bibfnamefont{R.}~\bibnamefont{Axelrod}} \bibnamefont{and}
  \bibinfo{author}{\bibfnamefont{W.}~\bibnamefont{Hamilton}},
  \bibinfo{journal}{Science} \textbf{\bibinfo{volume}{211}},
  \bibinfo{pages}{1390} (\bibinfo{year}{1981}).

\bibitem[{\citenamefont{Sigmund and Hofbauer}(1998)}]{Sigmund1998}
\bibinfo{author}{\bibfnamefont{J.~H.~K.} \bibnamefont{Sigmund}}
  \bibnamefont{and} \bibinfo{author}{\bibfnamefont{J.}~\bibnamefont{Hofbauer}},
  \emph{\bibinfo{title}{Evolutionary games and population dynamics}}
  (\bibinfo{publisher}{Cambridge Univ. Press, Cambridge},
  \bibinfo{year}{1998}).

\bibitem[{\citenamefont{Dugatkin and Reeve}(2000)}]{Dugatkin2000}
\bibinfo{author}{\bibfnamefont{L.~A.} \bibnamefont{Dugatkin}} \bibnamefont{and}
  \bibinfo{author}{\bibfnamefont{H.~K.} \bibnamefont{Reeve}},
  \emph{\bibinfo{title}{Game theory and animal behavior}}
  (\bibinfo{publisher}{Oxford Univ. Press, Oxford}, \bibinfo{year}{2000}).

\bibitem[{\citenamefont{Nowak}(2006)}]{Nowak2006}
\bibinfo{author}{\bibfnamefont{M.~A.} \bibnamefont{Nowak}},
  \bibinfo{journal}{Science} \textbf{\bibinfo{volume}{314}},
  \bibinfo{pages}{1560} (\bibinfo{year}{2006}).

\bibitem[{\citenamefont{Stewart et~al.}(2016)\citenamefont{Stewart, Parsons,
  and Plotkin}}]{Stewart2016}
\bibinfo{author}{\bibfnamefont{A.~J.} \bibnamefont{Stewart}},
  \bibinfo{author}{\bibfnamefont{T.~L.} \bibnamefont{Parsons}},
  \bibnamefont{and} \bibinfo{author}{\bibfnamefont{J.~B.}
  \bibnamefont{Plotkin}}, \bibinfo{journal}{Proc. Nat. Acad. Sci.}
  \textbf{\bibinfo{volume}{113}}, \bibinfo{pages}{E7003}
  (\bibinfo{year}{2016}).

\bibitem[{\citenamefont{Sigmund}(2010)}]{Sigmund2010}
\bibinfo{author}{\bibfnamefont{K.}~\bibnamefont{Sigmund}},
  \emph{\bibinfo{title}{The calculus of selfishness}}
  (\bibinfo{publisher}{Princeton University Press, Princeton},
  \bibinfo{year}{2010}).

\bibitem[{\citenamefont{Nowak and Sigmund}(1990)}]{Nowak1990}
\bibinfo{author}{\bibfnamefont{M.}~\bibnamefont{Nowak}} \bibnamefont{and}
  \bibinfo{author}{\bibfnamefont{K.}~\bibnamefont{Sigmund}},
  \bibinfo{journal}{Acta Appl. Matt.} \textbf{\bibinfo{volume}{20}},
  \bibinfo{pages}{247} (\bibinfo{year}{1990}).

\bibitem[{\citenamefont{Hauert and Schuster}(1997)}]{Hauert1997}
\bibinfo{author}{\bibfnamefont{C.}~\bibnamefont{Hauert}} \bibnamefont{and}
  \bibinfo{author}{\bibfnamefont{H.~G.} \bibnamefont{Schuster}},
  \bibinfo{journal}{Proc. R. Soc. B} \textbf{\bibinfo{volume}{264}},
  \bibinfo{pages}{513} (\bibinfo{year}{1997}).

\bibitem[{\citenamefont{Hilbe et~al.}(2017)\citenamefont{Hilbe,
  Martinez-Vaquero, Chatterjee, and Nowak}}]{Hilbe2017}
\bibinfo{author}{\bibfnamefont{C.}~\bibnamefont{Hilbe}},
  \bibinfo{author}{\bibfnamefont{L.~A.} \bibnamefont{Martinez-Vaquero}},
  \bibinfo{author}{\bibfnamefont{K.}~\bibnamefont{Chatterjee}},
  \bibnamefont{and} \bibinfo{author}{\bibfnamefont{M.~A.} \bibnamefont{Nowak}},
  \bibinfo{journal}{Proc. Nat. Acad. Sci.} \textbf{\bibinfo{volume}{114}},
  \bibinfo{pages}{4715} (\bibinfo{year}{2017}).

\bibitem[{\citenamefont{Boerlijst et~al.}(1997)\citenamefont{Boerlijst, Nowak,
  and Sigmund}}]{Boerlijst1997}
\bibinfo{author}{\bibfnamefont{M.~C.} \bibnamefont{Boerlijst}},
  \bibinfo{author}{\bibfnamefont{M.~A.} \bibnamefont{Nowak}}, \bibnamefont{and}
  \bibinfo{author}{\bibfnamefont{K.}~\bibnamefont{Sigmund}},
  \bibinfo{journal}{The Amer. Math. Month.} \textbf{\bibinfo{volume}{104}},
  \bibinfo{pages}{303} (\bibinfo{year}{1997}).

\bibitem[{\citenamefont{Press and Dyson}(2012)}]{Press2012}
\bibinfo{author}{\bibfnamefont{W.~H.} \bibnamefont{Press}} \bibnamefont{and}
  \bibinfo{author}{\bibfnamefont{F.~J.} \bibnamefont{Dyson}},
  \bibinfo{journal}{Proc. Nat. Acad. Sci.} \textbf{\bibinfo{volume}{109}},
  \bibinfo{pages}{10409} (\bibinfo{year}{2012}).

\bibitem[{\citenamefont{Adami and Hintze}(2013)}]{Adami2013}
\bibinfo{author}{\bibfnamefont{C.}~\bibnamefont{Adami}} \bibnamefont{and}
  \bibinfo{author}{\bibfnamefont{A.}~\bibnamefont{Hintze}},
  \bibinfo{journal}{Nat. Comm.} \textbf{\bibinfo{volume}{4}},
  \bibinfo{pages}{2193} (\bibinfo{year}{2013}).

\bibitem[{\citenamefont{Stewart and Plotkin}(2013)}]{Stewart2013}
\bibinfo{author}{\bibfnamefont{A.~J.} \bibnamefont{Stewart}} \bibnamefont{and}
  \bibinfo{author}{\bibfnamefont{J.~B.} \bibnamefont{Plotkin}},
  \bibinfo{journal}{Proc. Nat. Acad. Sci.} \textbf{\bibinfo{volume}{110}},
  \bibinfo{pages}{15348} (\bibinfo{year}{2013}).

\bibitem[{\citenamefont{Hilbe et~al.}(2015)\citenamefont{Hilbe, Traulsen, and
  Sigmund}}]{Hilbe2015}
\bibinfo{author}{\bibfnamefont{C.}~\bibnamefont{Hilbe}},
  \bibinfo{author}{\bibfnamefont{A.}~\bibnamefont{Traulsen}}, \bibnamefont{and}
  \bibinfo{author}{\bibfnamefont{K.}~\bibnamefont{Sigmund}},
  \bibinfo{journal}{Games and Econ. Behav.} \textbf{\bibinfo{volume}{92}},
  \bibinfo{pages}{41} (\bibinfo{year}{2015}).

\bibitem[{\citenamefont{Adami et~al.}(2016)\citenamefont{Adami, Schossau, and
  Hintze}}]{Adami2016}
\bibinfo{author}{\bibfnamefont{C.}~\bibnamefont{Adami}},
  \bibinfo{author}{\bibfnamefont{J.}~\bibnamefont{Schossau}}, \bibnamefont{and}
  \bibinfo{author}{\bibfnamefont{A.}~\bibnamefont{Hintze}},
  \bibinfo{journal}{Phys. Life Rev.} \textbf{\bibinfo{volume}{19}},
  \bibinfo{pages}{1} (\bibinfo{year}{2016}).

\bibitem[{\citenamefont{Milinski}(1987)}]{Milinski1987}
\bibinfo{author}{\bibfnamefont{M.}~\bibnamefont{Milinski}},
  \bibinfo{journal}{Nature} \textbf{\bibinfo{volume}{325}},
  \bibinfo{pages}{433} (\bibinfo{year}{1987}).

\bibitem[{\citenamefont{Lee et~al.}(2015)\citenamefont{Lee, Harper, and
  Fryer}}]{Lee2015}
\bibinfo{author}{\bibfnamefont{C.}~\bibnamefont{Lee}},
  \bibinfo{author}{\bibfnamefont{M.}~\bibnamefont{Harper}}, \bibnamefont{and}
  \bibinfo{author}{\bibfnamefont{D.}~\bibnamefont{Fryer}},
  \bibinfo{journal}{Plos ONE} \textbf{\bibinfo{volume}{10}}
  (\bibinfo{year}{2015}).

\bibitem[{\citenamefont{Austad}(1983)}]{Austad1983}
\bibinfo{author}{\bibfnamefont{S.~N.} \bibnamefont{Austad}},
  \bibinfo{journal}{Anim. Behav.} \textbf{\bibinfo{volume}{31}},
  \bibinfo{pages}{59} (\bibinfo{year}{1983}).

\bibitem[{\citenamefont{Smith and Price}(1973)}]{SMITH1973}
\bibinfo{author}{\bibfnamefont{J.}~\bibnamefont{Smith}} \bibnamefont{and}
  \bibinfo{author}{\bibfnamefont{G.}~\bibnamefont{Price}},
  \bibinfo{journal}{Nature} \textbf{\bibinfo{volume}{246}}, \bibinfo{pages}{15}
  (\bibinfo{year}{1973}).

\bibitem[{\citenamefont{Marden and Waage}(1990)}]{Marden1990}
\bibinfo{author}{\bibfnamefont{J.}~\bibnamefont{Marden}} \bibnamefont{and}
  \bibinfo{author}{\bibfnamefont{J.}~\bibnamefont{Waage}},
  \bibinfo{journal}{Anim. Behav.} \textbf{\bibinfo{volume}{39}},
  \bibinfo{pages}{954} (\bibinfo{year}{1990}).

\bibitem[{\citenamefont{Mesterton-Gibbons
  et~al.}(1996)\citenamefont{Mesterton-Gibbons, Marden, and
  Dugatkin}}]{MestertonGibbons1996}
\bibinfo{author}{\bibfnamefont{M.}~\bibnamefont{Mesterton-Gibbons}},
  \bibinfo{author}{\bibfnamefont{J.~H.} \bibnamefont{Marden}},
  \bibnamefont{and} \bibinfo{author}{\bibfnamefont{L.~A.}
  \bibnamefont{Dugatkin}}, \bibinfo{journal}{J.Theor. Biol.}
  \textbf{\bibinfo{volume}{181}}, \bibinfo{pages}{65} (\bibinfo{year}{1996}).

\bibitem[{\citenamefont{Prenter et~al.}(2006)\citenamefont{Prenter, Elwood, and
  Taylor}}]{Prenter2006}
\bibinfo{author}{\bibfnamefont{J.}~\bibnamefont{Prenter}},
  \bibinfo{author}{\bibfnamefont{R.~W.} \bibnamefont{Elwood}},
  \bibnamefont{and} \bibinfo{author}{\bibfnamefont{P.~W.}
  \bibnamefont{Taylor}}, \bibinfo{journal}{Anim. Behav.}
  \textbf{\bibinfo{volume}{72}}, \bibinfo{pages}{861} (\bibinfo{year}{2006}).

\bibitem[{\citenamefont{Dietemann et~al.}(2008)\citenamefont{Dietemann, Zheng,
  Hepburn, Hepburn, Jin, Crewe, Radloff, Hu, and Pirk}}]{Dietemann2008}
\bibinfo{author}{\bibfnamefont{V.}~\bibnamefont{Dietemann}},
  \bibinfo{author}{\bibfnamefont{H.-Q.} \bibnamefont{Zheng}},
  \bibinfo{author}{\bibfnamefont{C.}~\bibnamefont{Hepburn}},
  \bibinfo{author}{\bibfnamefont{H.~R.} \bibnamefont{Hepburn}},
  \bibinfo{author}{\bibfnamefont{S.-H.} \bibnamefont{Jin}},
  \bibinfo{author}{\bibfnamefont{R.~M.} \bibnamefont{Crewe}},
  \bibinfo{author}{\bibfnamefont{S.~E.} \bibnamefont{Radloff}},
  \bibinfo{author}{\bibfnamefont{F.-L.} \bibnamefont{Hu}}, \bibnamefont{and}
  \bibinfo{author}{\bibfnamefont{C.~W.~W.} \bibnamefont{Pirk}},
  \bibinfo{journal}{Plos One} \textbf{\bibinfo{volume}{3}}
  (\bibinfo{year}{2008}).

\bibitem[{\citenamefont{Keil and Watson}(2010)}]{Keil2010}
\bibinfo{author}{\bibfnamefont{P.~L.} \bibnamefont{Keil}} \bibnamefont{and}
  \bibinfo{author}{\bibfnamefont{P.~J.} \bibnamefont{Watson}},
  \bibinfo{journal}{Anim. Behav.} \textbf{\bibinfo{volume}{80}},
  \bibinfo{pages}{809} (\bibinfo{year}{2010}).

\bibitem[{\citenamefont{Percival and Moore}(2010)}]{Percival2010}
\bibinfo{author}{\bibfnamefont{D.~T.} \bibnamefont{Percival}} \bibnamefont{and}
  \bibinfo{author}{\bibfnamefont{P.~A.} \bibnamefont{Moore}},
  \bibinfo{journal}{Behav.} \textbf{\bibinfo{volume}{147}},
  \bibinfo{pages}{103} (\bibinfo{year}{2010}).

\bibitem[{\citenamefont{Martinez-Cotrina
  et~al.}(2014)\citenamefont{Martinez-Cotrina, Bohorquez-Alonso, and
  Molina-Borja}}]{Martinez-Cotrina2014}
\bibinfo{author}{\bibfnamefont{J.}~\bibnamefont{Martinez-Cotrina}},
  \bibinfo{author}{\bibfnamefont{M.~L.} \bibnamefont{Bohorquez-Alonso}},
  \bibnamefont{and}
  \bibinfo{author}{\bibfnamefont{M.}~\bibnamefont{Molina-Borja}},
  \bibinfo{journal}{Behav.} \textbf{\bibinfo{volume}{151}},
  \bibinfo{pages}{1535} (\bibinfo{year}{2014}).

\bibitem[{\citenamefont{Tsai et~al.}(2014)\citenamefont{Tsai, Barrows, and
  Weiss}}]{Tsai2014}
\bibinfo{author}{\bibfnamefont{Y.-J.~J.} \bibnamefont{Tsai}},
  \bibinfo{author}{\bibfnamefont{E.~M.} \bibnamefont{Barrows}},
  \bibnamefont{and} \bibinfo{author}{\bibfnamefont{M.~R.} \bibnamefont{Weiss}},
  \bibinfo{journal}{Ethol.} \textbf{\bibinfo{volume}{120}},
  \bibinfo{pages}{816} (\bibinfo{year}{2014}).

\bibitem[{\citenamefont{Enquist and Leimar}(1983)}]{Enquist1983}
\bibinfo{author}{\bibfnamefont{M.}~\bibnamefont{Enquist}} \bibnamefont{and}
  \bibinfo{author}{\bibfnamefont{O.}~\bibnamefont{Leimar}},
  \bibinfo{journal}{J. Theor. Biol.} \textbf{\bibinfo{volume}{102}},
  \bibinfo{pages}{387} (\bibinfo{year}{1983}).

\bibitem[{\citenamefont{Elwood and Arnott}(2012)}]{Elwood2012}
\bibinfo{author}{\bibfnamefont{R.~W.} \bibnamefont{Elwood}} \bibnamefont{and}
  \bibinfo{author}{\bibfnamefont{G.}~\bibnamefont{Arnott}},
  \bibinfo{journal}{Anim. Behav.} \textbf{\bibinfo{volume}{84}},
  \bibinfo{pages}{1095} (\bibinfo{year}{2012}).

\bibitem[{\citenamefont{Rillich et~al.}(2007)\citenamefont{Rillich,
  Schildberger, and Stevenson}}]{Rillich2007}
\bibinfo{author}{\bibfnamefont{J.}~\bibnamefont{Rillich}},
  \bibinfo{author}{\bibfnamefont{K.}~\bibnamefont{Schildberger}},
  \bibnamefont{and} \bibinfo{author}{\bibfnamefont{P.~A.}
  \bibnamefont{Stevenson}}, \bibinfo{journal}{Anim. Behav.}
  \textbf{\bibinfo{volume}{74}}, \bibinfo{pages}{823} (\bibinfo{year}{2007}).

\bibitem[{\citenamefont{Elias et~al.}(2008)\citenamefont{Elias, Kasumovic,
  Punzalan, Andrade, and Mason}}]{Elias2008}
\bibinfo{author}{\bibfnamefont{D.~O.} \bibnamefont{Elias}},
  \bibinfo{author}{\bibfnamefont{M.~M.} \bibnamefont{Kasumovic}},
  \bibinfo{author}{\bibfnamefont{D.}~\bibnamefont{Punzalan}},
  \bibinfo{author}{\bibfnamefont{M.~C.~B.} \bibnamefont{Andrade}},
  \bibnamefont{and} \bibinfo{author}{\bibfnamefont{A.~C.} \bibnamefont{Mason}},
  \bibinfo{journal}{Anim. Behav.} \textbf{\bibinfo{volume}{76}},
  \bibinfo{pages}{901} (\bibinfo{year}{2008}).

\bibitem[{\citenamefont{Mesterton-Gibbons and
  Heap}(2014)}]{Mesterton-Gibbons2014a}
\bibinfo{author}{\bibfnamefont{M.}~\bibnamefont{Mesterton-Gibbons}}
  \bibnamefont{and} \bibinfo{author}{\bibfnamefont{S.~M.} \bibnamefont{Heap}},
  \bibinfo{journal}{Amer. Natur.} \textbf{\bibinfo{volume}{183}},
  \bibinfo{pages}{199} (\bibinfo{year}{2014}).

\bibitem[{\citenamefont{Guillermo-Ferreira
  et~al.}(2015)\citenamefont{Guillermo-Ferreira, Gorb, Appel, Kovalev, and
  Bispo}}]{Guillermo-Ferreira2015}
\bibinfo{author}{\bibfnamefont{R.}~\bibnamefont{Guillermo-Ferreira}},
  \bibinfo{author}{\bibfnamefont{S.~N.} \bibnamefont{Gorb}},
  \bibinfo{author}{\bibfnamefont{E.}~\bibnamefont{Appel}},
  \bibinfo{author}{\bibfnamefont{A.}~\bibnamefont{Kovalev}}, \bibnamefont{and}
  \bibinfo{author}{\bibfnamefont{P.~C.} \bibnamefont{Bispo}},
  \bibinfo{journal}{Sci. of Nat.} \textbf{\bibinfo{volume}{102}}
  (\bibinfo{year}{2015}).

\bibitem[{\citenamefont{Benitez et~al.}(2017)\citenamefont{Benitez, Pappano,
  Beehner, and Bergman}}]{Benitez2017}
\bibinfo{author}{\bibfnamefont{M.~E.} \bibnamefont{Benitez}},
  \bibinfo{author}{\bibfnamefont{D.~J.} \bibnamefont{Pappano}},
  \bibinfo{author}{\bibfnamefont{J.~C.} \bibnamefont{Beehner}},
  \bibnamefont{and} \bibinfo{author}{\bibfnamefont{T.~J.}
  \bibnamefont{Bergman}}, \bibinfo{journal}{Sci. Rep.}
  \textbf{\bibinfo{volume}{7}} (\bibinfo{year}{2017}).

\bibitem[{\citenamefont{Baek et~al.}(2016)\citenamefont{Baek, Jeong, Hilbe, and
  Nowak}}]{Baek2016}
\bibinfo{author}{\bibfnamefont{S.~K.} \bibnamefont{Baek}},
  \bibinfo{author}{\bibfnamefont{H.~C.} \bibnamefont{Jeong}},
  \bibinfo{author}{\bibfnamefont{C.}~\bibnamefont{Hilbe}}, \bibnamefont{and}
  \bibinfo{author}{\bibfnamefont{M.~A.} \bibnamefont{Nowak}},
  \bibinfo{journal}{Sci. Rep.} \textbf{\bibinfo{volume}{6}}
  (\bibinfo{year}{2016}).

\bibitem[{\citenamefont{Tkacik and Bialek}(2016)}]{Tkacik2016}
\bibinfo{author}{\bibfnamefont{G.}~\bibnamefont{Tkacik}} \bibnamefont{and}
  \bibinfo{author}{\bibfnamefont{W.}~\bibnamefont{Bialek}}, in
  \emph{\bibinfo{booktitle}{Ann. Rev. of Cond. Mat. Phys.}}, edited by
  \bibinfo{editor}{\bibfnamefont{M.}~\bibnamefont{Marchetti}} \bibnamefont{and}
  \bibinfo{editor}{\bibfnamefont{S.}~\bibnamefont{Sachdev}}
  (\bibinfo{year}{2016}), vol.~\bibinfo{volume}{7}, pp.
  \bibinfo{pages}{89--117}.

\bibitem[{\citenamefont{Kussell and Leibler}(2005)}]{Kussell2005}
\bibinfo{author}{\bibfnamefont{E.}~\bibnamefont{Kussell}} \bibnamefont{and}
  \bibinfo{author}{\bibfnamefont{S.}~\bibnamefont{Leibler}},
  \bibinfo{journal}{Science} \textbf{\bibinfo{volume}{309}},
  \bibinfo{pages}{2075} (\bibinfo{year}{2005}).

\bibitem[{\citenamefont{Donaldson-Matasci
  et~al.}(2010)\citenamefont{Donaldson-Matasci, Bergstrom, and
  Lachmann}}]{Donaldson-Matasci2010}
\bibinfo{author}{\bibfnamefont{M.~C.} \bibnamefont{Donaldson-Matasci}},
  \bibinfo{author}{\bibfnamefont{C.~T.} \bibnamefont{Bergstrom}},
  \bibnamefont{and} \bibinfo{author}{\bibfnamefont{M.}~\bibnamefont{Lachmann}},
  \bibinfo{journal}{Oikos} \textbf{\bibinfo{volume}{119}}, \bibinfo{pages}{219}
  (\bibinfo{year}{2010}).

\bibitem[{\citenamefont{Schreiber}(2000)}]{Schreiber2000}
\bibinfo{author}{\bibfnamefont{T.}~\bibnamefont{Schreiber}},
  \bibinfo{journal}{Phys. Rev. Lett.} \textbf{\bibinfo{volume}{85}},
  \bibinfo{pages}{461} (\bibinfo{year}{2000}).

\bibitem[{\citenamefont{Parker and Rubinstein}(1981)}]{PARKER1981}
\bibinfo{author}{\bibfnamefont{G.}~\bibnamefont{Parker}} \bibnamefont{and}
  \bibinfo{author}{\bibfnamefont{D.}~\bibnamefont{Rubinstein}},
  \bibinfo{journal}{Anim. Behav.} \textbf{\bibinfo{volume}{29}},
  \bibinfo{pages}{221} (\bibinfo{year}{1981}).

\bibitem[{\citenamefont{Parker}(1974)}]{Parker1974}
\bibinfo{author}{\bibfnamefont{G.~A.} \bibnamefont{Parker}},
  \bibinfo{journal}{J. Theor. Biol.} \textbf{\bibinfo{volume}{47}},
  \bibinfo{pages}{223} (\bibinfo{year}{1974}).

\bibitem[{\citenamefont{Hammerstein and Parker}(1982)}]{HAMMERSTEIN1982}
\bibinfo{author}{\bibfnamefont{P.}~\bibnamefont{Hammerstein}} \bibnamefont{and}
  \bibinfo{author}{\bibfnamefont{G.~A.} \bibnamefont{Parker}},
  \bibinfo{journal}{J. Theor. Biol.} \textbf{\bibinfo{volume}{96}},
  \bibinfo{pages}{647} (\bibinfo{year}{1982}).

\bibitem[{\citenamefont{Bishop and Cannings}(1978)}]{Bishop1978}
\bibinfo{author}{\bibfnamefont{D.}~\bibnamefont{Bishop}} \bibnamefont{and}
  \bibinfo{author}{\bibfnamefont{C.}~\bibnamefont{Cannings}},
  \bibinfo{journal}{J. of Theor. Biol} \textbf{\bibinfo{volume}{70}},
  \bibinfo{pages}{85} (\bibinfo{year}{1978}).

\bibitem[{\citenamefont{Payne and Pagel}(1996)}]{Payne1996a}
\bibinfo{author}{\bibfnamefont{R.}~\bibnamefont{Payne}} \bibnamefont{and}
  \bibinfo{author}{\bibfnamefont{M.}~\bibnamefont{Pagel}}, \bibinfo{journal}{J.
  of Theor. Biol.} \textbf{\bibinfo{volume}{183}}, \bibinfo{pages}{185}
  (\bibinfo{year}{1996}).

\bibitem[{\citenamefont{Stuart-Fox}(2006)}]{Stuart-Fox2006}
\bibinfo{author}{\bibfnamefont{D.}~\bibnamefont{Stuart-Fox}},
  \bibinfo{journal}{Proc. of the R. Soc. B} \textbf{\bibinfo{volume}{273}},
  \bibinfo{pages}{1555} (\bibinfo{year}{2006}).

\bibitem[{\citenamefont{Takeuchi et~al.}(2016)\citenamefont{Takeuchi, Yabuta,
  and Tsubaki}}]{Takeuchi2016}
\bibinfo{author}{\bibfnamefont{T.}~\bibnamefont{Takeuchi}},
  \bibinfo{author}{\bibfnamefont{S.}~\bibnamefont{Yabuta}}, \bibnamefont{and}
  \bibinfo{author}{\bibfnamefont{Y.}~\bibnamefont{Tsubaki}},
  \bibinfo{journal}{Biol. J. Lin. Soc.} \textbf{\bibinfo{volume}{118}},
  \bibinfo{pages}{970} (\bibinfo{year}{2016}).

\bibitem[{\citenamefont{Kim and Lee}(2014)}]{Kim2014}
\bibinfo{author}{\bibfnamefont{K.}~\bibnamefont{Kim}} \bibnamefont{and}
  \bibinfo{author}{\bibfnamefont{F.~Z.~X.} \bibnamefont{Lee}},
  \bibinfo{journal}{Amer. Econ. J.} \textbf{\bibinfo{volume}{6}},
  \bibinfo{pages}{37} (\bibinfo{year}{2014}).

\bibitem[{\citenamefont{Kirman}(1993)}]{Kirman1993}
\bibinfo{author}{\bibfnamefont{A.}~\bibnamefont{Kirman}},
  \bibinfo{journal}{Quat. J. Econ.} \textbf{\bibinfo{volume}{108}},
  \bibinfo{pages}{137} (\bibinfo{year}{1993}).

\bibitem[{\citenamefont{Doebeli and Hauert}(2005)}]{Doebeli2005}
\bibinfo{author}{\bibfnamefont{M.}~\bibnamefont{Doebeli}} \bibnamefont{and}
  \bibinfo{author}{\bibfnamefont{C.}~\bibnamefont{Hauert}},
  \bibinfo{journal}{Ecol. Lett.} \textbf{\bibinfo{volume}{8}},
  \bibinfo{pages}{748} (\bibinfo{year}{2005}).

\bibitem[{\citenamefont{Gore et~al.}(2009)\citenamefont{Gore, Youk, and van
  Oudenaarden}}]{Gore2009}
\bibinfo{author}{\bibfnamefont{J.}~\bibnamefont{Gore}},
  \bibinfo{author}{\bibfnamefont{H.}~\bibnamefont{Youk}}, \bibnamefont{and}
  \bibinfo{author}{\bibfnamefont{A.}~\bibnamefont{van Oudenaarden}},
  \bibinfo{journal}{Nature} \textbf{\bibinfo{volume}{459}},
  \bibinfo{pages}{253} (\bibinfo{year}{2009}).

\bibitem[{\citenamefont{Perc and Szolnoki}(2010)}]{Perc2010}
\bibinfo{author}{\bibfnamefont{M.}~\bibnamefont{Perc}} \bibnamefont{and}
  \bibinfo{author}{\bibfnamefont{A.}~\bibnamefont{Szolnoki}},
  \bibinfo{journal}{BioSys.} \textbf{\bibinfo{volume}{99}},
  \bibinfo{pages}{109} (\bibinfo{year}{2010}).

\bibitem[{\citenamefont{Leimar et~al.}(1991)\citenamefont{Leimar, Austad, and
  Enquist}}]{Leimar1991}
\bibinfo{author}{\bibfnamefont{O.}~\bibnamefont{Leimar}},
  \bibinfo{author}{\bibfnamefont{S.}~\bibnamefont{Austad}}, \bibnamefont{and}
  \bibinfo{author}{\bibfnamefont{M.}~\bibnamefont{Enquist}},
  \bibinfo{journal}{Evol.} \textbf{\bibinfo{volume}{45}}, \bibinfo{pages}{862}
  (\bibinfo{year}{1991}).

\end{thebibliography}

\section{Appendix A: Outcome Probabilities \(\vec{\Omega}\) of Equalizer Strategies}
\label{sec:org13e5bf7}
\label{SEC:appit} 

To calculate probabilities of the outcomes \(\vec{\Omega}_{M_{1}}(\vec{p}_{M_{1}},\vec{q}_{M_{1}})=(\Omega_{CC},\Omega_{CD},\Omega_{DC},\Omega_{DD})\) between two memory-one strategies \(\vec{p}_{M_{1}}=(p_{CC},p_{CD},p_{DC},p_{DD})\) and \(\vec{q}_{M_{1}}=(q_{CC},q_{CD},q_{DC},q_{DD})\), one finds the stationary eigenvector: 
\begin{eqnarray}
\label{zd543}
 M \vec{\Omega}=\vec{\Omega},
\end{eqnarray}
of Markov matrix\cite{Hauert1997} \(M\): 
\onecolumngrid 
\begin{eqnarray}
\label{M875483}
&&M (\vec{p}_{M_{1}},\vec{q}_{M_{1}})=\\
&&\left (\begin{array}{cccc}
p_{CC}q_{CC}&p_{CD}q_{DC} & p_{DC}q_{CD} & p_{DD}q_{DD} \\
p_{CC}(1-q_{CC}) & p_{CD}(1-q_{DC}) & p_{DC}(1-q_{CD}) & p_{DD}(1-q_{DD}) \\
(1-p_{CC})q_{CC} & (1-p_{CD})q_{DC} & (1-p_{DC})q_{CD} & (1-p_{DD})q_{DD} \\
(1-p_{CC})(1-q_{CC}) & (1-p_{CD})(1-q_{DC}) & (1-p_{DC})(1-q_{CD}) & (1-p_{DD})(1-q_{DD}) \\
\end{array}\right ), \nonumber 
\end{eqnarray}
\twocolumngrid 
between probabilities of the outcomes of the subsequent rounds of interaction \(\Omega_{N}=M\Omega_{N-1}\). 

The gain \(G(\vec{p},\vec{q},\vec{W}_{p})\) of strategy \(\vec{p}\) against strategy \(\vec{q}\) is: 
\begin{eqnarray}
\label{eq:67} G(\vec{p},\vec{q},\vec{W}_{p})=\vec{\Omega}(\vec{p},\vec{q})\cdot\vec{W},
\end{eqnarray}
where \(\vec{W}=(W_{CC},W_{CD},W_{DC},W_{DD})\) are the payoffs for interactions \((CC,CD,DC,DD)\) correspondingly, see (\ref{eq:1}). 

Press and Dyson derived a general formula \cite{Press2012} for the gains of memory-one strategies \(\vec{p}\) and \(\vec{q}\): 
\begin{eqnarray}
\label{eq:40}
&&\alpha G(\vec{p},\vec{q})+\beta G(\vec{q},\vec{p})+\gamma=\\\nonumber
&&\frac{\det\left (M_{G}\left(\vec{p},\vec{q},\alpha \vec{W}_{p}+\beta \vec{W}_{q}+\gamma\vec{1} \right )\right )}{\det\left(M_{G}\left(\vec{p},\vec{q},\vec{1} \right )\right )},
\end{eqnarray}
where: 
\begin{eqnarray}
\label{M7477}
&&M_{G}(\vec{p},\vec{q},\vec{W}_{p})=\\
&&\left [\begin{array}{cccc}
-1+p_{CC}q_{CC} & -1+p_{CC} & -1+q_{CC} & W_{CC} \\
p_{CD}q_{DC} & -1+p_{CD} & q_{DD} & W_{CD} \\
p_{DC}q_{CD} & p_{DC} & -1+q_{CD} & W_{DC} \\
p_{DD}q_{DD} & p_{DD} & q_{DD} & W_{DD} \\
\end{array}\right ], \nonumber 
\end{eqnarray}
and \(\alpha\) ,\(\beta\) and \(\gamma\) are arbitrary numbers. 

Zero-determinant strategies \(\vec{p}\) fit: 
\begin{eqnarray}
\label{eq:42}
\vec{p}=\alpha \vec{W}_{p}+\beta \vec{W}_{q}+\gamma\vec{1}.
\end{eqnarray}
In this case, determinant on the right side of the expression (\ref{eq:40}) vanish resulting in a linear relation between the gains of opponents \(\vec{p}\) and \(\vec{q}\): 
\begin{eqnarray}
\label{eq:45}
\alpha G(\vec{p},\vec{q})+\beta G(\vec{q},\vec{p})+\gamma=0.
\end{eqnarray} 
Thus, a player can impose specific constraints on the gain of its opponent by choosing the values of \(\alpha\), \(\beta\) and \(\gamma\) in (\ref{eq:45}), and by choosing its strategy \(\vec{p}\) to fit (\ref{eq:42}). 

Equalizer strategies (\ref{eq:15986}) correspond to \(\alpha=0\) in (\ref{eq:45}). In this case, strategy \(p\) imposes payoff \(-\gamma/\beta\) on any strategy \(\vec{q}\). Equalizer strategies possess two independent variables \((p_{CC},p_{DD})\). The probabilities \(p_{CD}\) and \(p_{DC}\) (\ref{eq:15986}) follow from (\ref{eq:42}). 

Constraints on possible values of the probabilities in a memory-one strategy \(0\leq p_{CC},p_{CD},p_{DC},p_{DD}\leq 1\) impose boundaries on possible equalizer strategies \((p_{CC},p_{DD})\). 

In the case of conflict escalation \(0<S<1\) and \(S<T\), the boundary of the valid strategies has two linear segments on the left, see Figure \ref{Occeqv}. The lower segment is: 
\begin{eqnarray}
\label{eq:71}
p_{DD}=\frac{S(1-p_{CC})}{1-S},\;\;\frac{T-1}{T-S}\leq p_{CC}\leq 1,
\end{eqnarray}
and the upper segment is: 
\begin{eqnarray}
\label{eq:71754}
p_{DD}=\frac{1-T+T p_{CC}}{T-1},\;\;\frac{T-1}{T-S}\leq p_{CC}\leq \frac{2(T-1)}{T}.
\end{eqnarray}
The other boundaries of the valid strategies coincide with boundaries of the strategy space: \((\frac{2(T-1)}{T}\leq p_{CC}\leq 1,p_{DD}=1)\) and \((p_{CC}=1,0\leq p_{DD}\leq 1)\). 

In the case of prisoner's dilemma \(S<0\), the boundaries of the valid strategies are either of two types. First, if \(S>-T\), then there are two left linear segments. The lower segment is: 
\begin{eqnarray}
\label{eq:775}
p_{DD}=\frac{1-T+T p_{CC}}{T-1},\;\;\frac{T-1}{T}\leq p_{CC}\leq \frac{2(T-1)}{T-S},
\end{eqnarray}
and the upper segment is: 
\begin{eqnarray}
\label{eq:79654}
p_{DD}=\frac{-1-S+S p_{CC}}{S-1},\;\;\frac{2(T-1)}{T-S} \leq p_{CC}\leq 1.
\end{eqnarray}
Second, if \(S<-T\), then there is a single left segment: 
\begin{eqnarray}
\label{eq:9754}
p_{DD}=\frac{-1-S+S p_{CC}}{S-1},\;\;\frac{S+1}{S} \leq p_{CC}\leq 1,
\end{eqnarray}
Other boundaries, like in the case of the snowdrift game, go along the boundaries of the strategy space, see Figure \ref{Occeqv1}.

Following (\ref{zd543}), probabilities of interactions between two identical equalizer strategies \((p_{CC},p_{DD})\) are: 
\begin{eqnarray}
\label{eq:13}
&&\Omega_{CC}=p_{DD} (-2 (-1 + p_{CC}) S + \\\nonumber
&&p_{DD} (-2 + (3 - p_{CC} + p_{DD}) S) - (-1 + p_{CC} - p_{DD}) \\\nonumber
&&(p_{DD} + 2 (-1 + p_{CC} - p_{DD}) S) T)/D,\\\nonumber
&&\Omega_{CD}=\Omega_{DC}=((-1 + p_{CC}) (1 + p_{CC} - p_{DD}) p_{DD})/D,\\\nonumber
&&\Omega_{DD}=((-1 + p_{CC}) \times \\\nonumber
&&\left (-2 p_{CC} p_{DD} + 2 p_{DD}^2 - S + 2 p_{CC} S - p_{CC}^2 S - 3 p_{DD} S\right .\\\nonumber
&& + 3 p_{CC} p_{DD} S - 2 p_{DD}^2 S + (-1 + p_{CC} - p_{DD}) \times \\\nonumber
&&\left . (-1 + 2 p_{DD} - 2 (1 + p_{DD}) S + p_{CC} (-1 + 2 S)) T)\right )/D,
\end{eqnarray}
where denominator \(D\) is: 
\begin{eqnarray}
\label{eq:68}
&&D = (-1 + p_{CC} - p_{DD}) \times \\\nonumber
&&\left (-(-1 + p_{CC})^2 S + p_{DD} (2 - (4 - 2 p_{CC} + p_{DD}) S) + T + \right .\\\nonumber
&&(-2 p_{DD} + 2 S + (p_{CC} - p_{DD}) (p_{DD} - \\\nonumber
&&\left . 2 (2 + p_{DD}) S + p_{CC} (-1 + 2 S))) T\right ).
\end{eqnarray}
These expressions may be calculated for any values of \((p_{CC},p_{DD})\) but are valid only within the boundaries (\ref{eq:71}-\ref{eq:9754}). 

Outcome probabilities \(\Omega\) as a function of an equalizer strategy have discontinuity at point \((p_{CC},p_{DD})=(1,0)\). The limit depends on the path of approach: 
\begin{eqnarray}
\label{eq:6}
p_{DD}=k (-S + p_{CC} S)/(-1 + S),
\end{eqnarray}
that depends on parameter \(1\leq k\). If \(k=1\) then (\ref{eq:6}) corresponds to the lower boundary (\ref{eq:71}). The corresponding limits of the outcome probabilities are: 
\begin{eqnarray}
\label{eq:65}
&&\Omega_{CC}=\frac{k(k-1)S^{2}}{(1+(k-1)S)(kS+T)},\nonumber\\
&&\Omega_{CD}=\Omega_{DC}=\frac{kS}{(1+(k-1)S)(kS+T)},\nonumber\\
&&\Omega_{DD}=\frac{kS(T-1)+T-ST}{(1+(k-1)S)(kS+T)},
\end{eqnarray}
Only for \(k=1\) there exists a single strategy with \(\Omega_{CC}=0\), see Figure \ref{Occeqv}. This strategy exists for \(0\leq S\leq 1\) and \(S\leq T\): from condition \(\Omega_{CD}+\Omega_{DC}<1\) follows that \(S<T\), and from \(S/(1-S)>0\) (condition  for the boundary to exist) follows \(0<S<1\). The case \(k=0\) occurs in the case of prisoner’s dilemma and leads to degenerate \(\vec{\Omega}=(0,0,0,1)\) (\ref{eq:65}). 

\section{Appendix B: In the case of the Snowdrift game, equalizer strategies are unique in their ability to match outcome probabilities between an arbitrary strategy and its mutants}
\label{sec:org7e88ff2}
\label{SEC:appuniq} 

Let us show that, in the case of the snowdrift game, if (\ref{eq:64}) holds then \(p\) is an equalizer strategy. In this case, the linear expansion of outcome probabilities in a competition between a memory-one strategy and its near-mutants form a plane in the \(\Omega\) space. The plane is perpendicular to the payoffs vector \(W\).

First, following (\ref{eq:40}) with \(\beta=1,\alpha=\gamma=0\), any equalizer strategy fits:
\begin{eqnarray}
\label{eq:6986}
\left . \frac{\partial G(\vec{q},\vec{p})}{\partial q_{i}}\right |_{\vec{p}=\vec{q}}=0,
\end{eqnarray}
where index \(i\) takes the values CC,CD,DC, or DD. Second, assume that \(\vec{q}\) is not an equalizer strategy. Then \(\det\left (M_{G}\left(\vec{p},\vec{q}, \vec{W}_{q} \right )\right )\neq 0\) and applying the Jacobi formula for derivative of a determinant, one gets: 
\begin{eqnarray}
\label{eq:41}
 &&\frac{\partial G(\vec{q},\vec{p})}{\partial q_{i}}=\frac{\det\left (M_{G}\left(\vec{p},\vec{q}, \vec{W}_{q} \right )\right )}{\det\left(M_{G}\left(\vec{p},\vec{q},\vec{1} \right )\right )}\times\nonumber\\
 &&Tr\left [ \right . -\frac{M^{-1}_{G}\left(\vec{p},\vec{q},\vec{1} \right )}{\det\left(M_{G}\left(\vec{p},\vec{q},\vec{1} \right )\right )}\frac{\partial M_{G}\left(\vec{p},\vec{q},\vec{1} \right )}{\partial q_{i}}\left |_{p=q}+\right .\nonumber\\
 &&\left . \left . M^{-1}_{G}\left(\vec{p},\vec{q}, \vec{W}_{q} \right )\frac{\partial M_{G}\left(\vec{p},\vec{q}, \vec{W}_{q} \right )}{\partial q_{i}}\right |_{p=q}\right ].
\end{eqnarray}
Thus a non-equalizer strategy that fits (\ref{eq:6986}) should vanish the \(Tr\) operator in (\ref{eq:41}). 

The equations (\ref{eq:41}) constitute a system of cumbersome non-linear equations that, nevertheless, can be reduced to three functions \(p_{CC}(p_{CD},p_{DC},p_{DD}),T(p_{CD},p_{DC},p_{DD})\), and \(S(p_{CD},p_{DC},p_{DD})\). Then, these functions are scanned numerically for solutions that fit the conditions: 
\begin{eqnarray}
\label{eq:5}
 &&0<=p_{CC},p_{CD},p_{DC},p_{DD}<=1,\\\nonumber
 &&0\leq S\leq 1,
\end{eqnarray}
for the valid values of the probabilities and the payoffs of the snowdrift game. No solution fits condition (\ref{eq:5}) for the snowdrift game \(0<S<1\). For the payoffs other than the snowdrift game, e.g. prisoner's dilemma \(S<0\), there exist local solutions that fit (\ref{eq:5}). 

\section{Appendix C: Spiders}
\label{sec:orge5d83d2}
\label{SEC:appspid} 

S. Austad collected substantial data regarding combat between male bowl and doily spiders. The data includes statistics on fatal injuries, the number of eggs in a female nest, the value of the female nest for a winner and his less fortunate opponent (by second insemination), the average lifetime reproductive success, and the duration of the fights. The ratio of fights that end with a severe injury depends on the difference in the competitors' sizes: a smaller spider generally flees from a larger opponent. This work addresses only the fights between spiders of similar size.

Let us present the mating combat of bowl and doily spiders as a two-person game. Each spider decides between two roles: to flee \(C\) or to fight until death D. Under the assumption of a single fight per life, the payoffs \((R,S,T,P)\) are: 
\begin{eqnarray}
\label{eq:34}
&&R=\frac{V_{F}}{2}+V_{\substack{rest \\ life}},S=\delta V_{F}+V_{\substack{rest \\ life}},\nonumber\\
&&T=(1-\delta) V_{F}+V_{\substack{rest \\ life}},P=\frac{V_{F}}{2}+\frac{V_{\substack{rest \\ life}}}{2},
\end{eqnarray}
where \(V_{F}\) is the amount of unfertilized eggs in the female's nest, and \(V_{\substack{rest \\ life}}\) is an estimate of the amount of eggs to fertilize after the first fight. Payoff \(R\) corresponds to a \(CC\) fight where each player gets half of the nest and all the value of the future amount of eggs. In a \(CD\) competition, \(D\) gets \(T\) equal to almost the total value of the nest \((1-\delta )V_{F}\) together with the value of the future amount of eggs, while \(C\) gets \(S\) equal to \(\delta V_{F}\) by the second insemination together with the value of the future amount of eggs. Payoff \(P\) corresponds to \(DD\) fight with equal probability to die and get \(0\), or get the total value of the nest together with the value of the future amount of eggs. 

In the case of bowl and doily spiders, the value of payoffs (\ref{eq:34}) can be estimated from the data collected by S. Austad\cite{Austad1983}. The lifetime payoff of a spider without the fighting cost is \(V_{L}=16.2\) eggs. The average value of a female's nest is \(V_{F}=10\) eggs. Then, we can estimate that a spider that manages to flee uninjured from the first fight can expect \(V_{L}-V_{F}/2<V_{\substack{rest \\ life}}<V_{L}\). Thus, \(V_{\substack{rest \\ life}}\approx 13.5\) eggs. This value of the future amount of eggs is in agreement with the previous estimations\cite{Leimar1991}. The spider that flees from the competitor gets \(\delta\approx 5\%\) of the nest value by the second insemination. 

Substituting the values from the previous paragraph to (\ref{eq:34}) and applying a normalization procedure (\ref{eq:8}), one gets (\ref{eq:38}).

Statistics of the fight outcomes for bowl and doily spiders is estimated as (\ref{eq:20}). In the case of bowl and doily spiders\cite{Austad1983}, \(0.67\) of the fights between opponents of similar sizes result in the death of a participant. Thus \(\Omega_{DD}=0.67\), taking into account that the death of a competitor occurs only in a \(DD\) type competition, because both in \(CD\) and in \(DC\), the \(C\) competitor flees and, therefore, stays alive. Mutual cooperation \(CC\) never happens \(\Omega_{CC}=0\). Then result (\ref{eq:20}) follows because \(\Omega_{CC}+\Omega_{DC}+\Omega_{CD}+\Omega_{DD}=1\) and \(\Omega_{CD}=\Omega_{DC}\) due to symmetry. 

\section{Appendix D: Outcome Probabilities \(\vec{\Omega}\) of \(h_{K}=(P_{C|D},P_{C|C})\) Strategies}
\label{sec:org88978a2}
\label{SEC:appnonit} 

Consider multiple interactions of players \(i\) and \(j\) with the strategies \((P^{i}_{C|D},P^{i}_{C|C})\) and \((P^{j}_{C|D},P^{j}_{C|C})\) (\ref{eq:21}), respectively. If player \(i\) is the first to respond, then, after \(N\) rounds, the average probability \(P_{D}^{ji}\) of player \(j\) to be in state \(D\) is: 
\begin{eqnarray}
\label{eq:33} 
P^{ji}_{D}=\frac{1}{N}\sum_{k=1}^{N}\left [(1-P^{j}_{C|D})\delta_{S^{ij}(k),D}+(1-P^{j}_{C|C})\delta_{S^{ij}(k),C}\right ],
\end{eqnarray}
where \(S^{i}(k)\) is the state (\(C\) or \(D\)) of the player \(i\) against player \(j\) in round \(k\) and \(\delta\) is Kronecker delta (\(\delta_{ij}=1\) if \(i=j\) and \(\delta_{ij}=0\) if \(i\neq j\)). 

Following (\ref{eq:33}), the average probability \(P_{D}^{ij}\) of player \(i\) to be in state \(D\) is: 
\begin{eqnarray}
\label{eq:43}
\frac{1}{N}\sum_{k=1}^{N}\delta_{S^{ij}(k),D}=P^{ij}_{D},
\end{eqnarray} 
Under the assumption that a player possesses the same probability to be in state \(D\) while responding as the first or second, eqs. (\ref{eq:33}) and (\ref{eq:43}) become: 
\begin{eqnarray}
P^{ji}_{D}=(1-P^{j}_{C|D})P^{ij}_{D}+(1-P^{j}_{C|C})(1-P^{ij}_{D}),\nonumber \\
P^{ij}_{D}=(1-P^{i}_{C|D})P^{ji}_{D}+(1-P^{i}_{C|C})(1-P^{ji}_{D}), \label{SMgammajisys}%
\end{eqnarray}
The same equations (\ref{SMgammajisys}) can be justified by symmetry consideration without addressing multiple interactions. 

The system (\ref{SMgammajisys}) can be solved for \(P^{ij}_{D}\) and \(P^{ji}_{D}\): 
\begin{equation}
P^{ji}_{D}=\frac{(1-P^{j}_{C|C})-(1-P^{i}_{C|C})(P^{j}_{C|D}-P^{j}_{C|C})}{{1-(P^{i}_{C|D}-P^{i}_{C|C})(P^{j}_{C|D}-P^{j}_{C|C})}}. \label{SMgammaji}%
\end{equation}
and 
\begin{eqnarray}
\label{eq:23}
P^{ij}_{D} = \frac{1 - P^{i}_{C|C} - (1 - P^{j}_{C|C}) (P^{i}_{C|D} - P^{i}_{C|C})}{1- (P^{i}_{C|D} - P^{i}_{C|C}) (P^{j}_{C|D} - P^{j}_{C|C})},
\end{eqnarray}
where \((P^{i}_{C|D},P^{i}_{C|C})\) and \((P^{j}_{C|D},P^{j}_{C|C})\) are the strategies of the competitors. 

Here are the probabilities \((\Omega_{CC},\Omega_{CD},\Omega_{DC},\Omega_{DD})\) of interactions of types \((CC,CD,DC,DD)\) between players \(i\) and \(j\) under the condition that player \(i\) is the first, and player \(j\) is the second to respond: 
\begin{eqnarray}
\label{eq:37}
&&\Omega_{CC}= P^{j}_{C|C} (1 - P^{ij}_{D}),\Omega_{CD}= (1 - P^{j}_{C|C}) (1 - P^{ij}_{D}),\nonumber \\
&&\Omega_{DC}=P^{ij}_{D} P^{j}_{C|D}, \Omega_{DD}=P^{ij}_{D} (1-P^{j}_{C|D}),
\end{eqnarray}
where \(P_{D}^{ji}\) and \(P_{D}^{ij}\) are the probabilities of the players \(j\) and \(i\) to be in state \(D\), respectively. If player \(i\) is the second to respond, expressions (\ref{eq:37}) are valid with a swap of the indices \(i\leftrightarrow j\). 

The gains (\ref{eq:44}) of player \(i\) are: 
\begin{eqnarray}
\label{eq:25}
G^{ij}_{1} = P^{i}_{D} P^{j}_{C|D} T + (1 - P^{j}_{C|C}) (1 - P^{i}_{D}) S + P^{j}_{C|C} (1 - P^{i}_{D}),
\end{eqnarray}
if it is the first, and: 
\begin{eqnarray}
\label{eq:2523}
G^{ij}_{2} = P^{j}_{D} P^{i}_{C|D} S + (1 - P^{i}_{C|C}) (1 - P^{j}_{D}) T + P^{i}_{C|C} (1 - P^{j}_{D}),
\end{eqnarray}
if it is the second. Under the condition that a player possesses equal probabilities to be the second or the first, the total gain is: 
\begin{eqnarray}
\label{eq:26}
G^{ij} = \frac{G^{ij}_{1}+G^{ij}_{2}}{2},
\end{eqnarray}
where \(G^{ij}\) is a function of the strategies of both competitors. 

Adaptive dynamics formalism is used to find an evolutionarily stable strategy \((P_{C|D}^{ESS},P_{C|C}^{ESS})\). The population is presented as a point in space that moves with velocity \(\vec{V}=(V_{P_{C|D}},V_{P_{C|C}})\). The direction and absolute value of velocity correspond to the optimal gradient of a mutant's fitness: 
\begin{eqnarray}
\label{eq:27}
V_{P_{C|D}} &=& \left . \frac{\partial G^{ij}}{\partial P^{i}_{C|D}}\right |_{
\substack{P^{i}_{C|D}=P^{j}_{C|D} \\ P^{i}_{C|C}=P^{j}_{C|C} }},\nonumber\\
V_{P_{C|C}} &=& \left . \frac{\partial G^{ij}}{\partial P^{i}_{C|C}}\right |_{
\substack{P^{i}_{C|D}=P^{j}_{C|D} \\ P^{i}_{C|C}=P^{j}_{C|C} }},
\end{eqnarray}
when mutation steps are assumed to be small. The stable points of flow field \(\vec{V}\) correspond to the ESS states. It corresponds to condition (\ref{eq:16}) under the constraint of small mutation steps. 

Adaptive dynamics (\ref{eq:27}) converges to a steady homogeneous population of identical individuals \((P^{i}_{C|D},P^{i}_{C|C})=(P^{j}_{C|D},P^{j}_{C|C})\). In this population, the probability of each player to be at state \(D\) reduces to: 
\begin{eqnarray}
\label{eq:24}
P^{ii}_{D} = \frac{1 - P^{i}_{C|C}}{1 + P^{i}_{C|D} - P^{i}_{C|C}},
\end{eqnarray}
following (\ref{SMgammaji}) and (\ref{eq:23}). Abundance of state \(D\) can take any value \(0\leq P_{D}\leq 1\) as a function of a strategy \((P_{C|D},P_{C|C})\). 

In the case of the snowdrift game, adaptive dynamics in space \((P_{C|D},P_{C|C})\) is presented as a vector field in Fig. \ref{RepDynSD}. There are two convergence regions: (\ref{eq:29}) (marked by red circle) and the segment (marked by green dashed line): 
\begin{eqnarray}
\label{eq:52}
&P_{C|C}&=1,\nonumber\\,
0\leq &P_{C|D}&\leq P^{SD_{1}}_{C|D},
\end{eqnarray}
where: 
\begin{eqnarray}
\label{eq:545}
&& P^{SD_{1}}_{C|D}=\\
&&\frac{-1 + S + 3T - \sqrt{9 - 10S + S^{2} - 14T + 6ST + 9T^{2}}}{2(-1 + S + T)}.\nonumber
\end{eqnarray}
The first point is an ESS (\ref{eq:16}). The second segment (\ref{eq:52}) is not an ESS as defined by (\ref{eq:16}) because the gains are equal for the hosts and the mutants along the boundary \(P_{C|C}=1\). 

Construction of \(\vec{W}'\) (\ref{eq:60}) for strategy (\ref{eq:6}) results in (\ref{eq:4}). The corresponding:
\begin{widetext}
\begin{eqnarray}
\label{eq:10}
&&\pi_{1}=\\\nonumber
&&\left\{-\frac{(P_{C|C}-1) P_{C|C}}{2 (-P_{C|C}+P_{C|D}+1)^2},\frac{(P_{C|C}-1)
  \left(P_{C|C}^2-P_{C|C} P_{C|D}+P_{C|C}-2\right)}{2 (P_{C|C}-P_{C|D}+1)
  (-P_{C|C}+P_{C|D}+1)^2},\frac{(P_{C|C}-1) \left(P_{C|C}^2-P_{C|C} (P_{C|D}+1)+2
  P_{C|D}\right)}{2 (P_{C|C}-P_{C|D}+1) (-P_{C|C}+P_{C|D}+1)^2}\right\},
\end{eqnarray}
\end{widetext}
and 
\begin{widetext}
\begin{eqnarray}
\label{eq:49}
&&\pi_{2}=\\\nonumber
&&\left\{\frac{P_{C|D} (P_{C|D}+1)}{2 (-P_{C|C}+P_{C|D}+1)^2},\frac{P_{C|D}
  \left(-P_{C|C} (P_{C|D}+1)+P_{C|D}^2+1\right)}{2 (P_{C|C}-P_{C|D}+1)
  (-P_{C|C}+P_{C|D}+1)^2},\frac{P_{C|D} \left(-P_{C|C} P_{C|D}+P_{C|C}+P_{C|D}^2-2
  P_{C|D}-1\right)}{2 (P_{C|C}-P_{C|D}+1) (-P_{C|C}+P_{C|D}+1)^2}\right\},
\end{eqnarray}
\end{widetext}
follow substitution of (\ref{eq:37}) to (\ref{eq:47}). 

ESS (\ref{eq:6}) is unique for the snowdrift game. Adaptive dynamics in the case of prisoner's dilemma possesses two convergence regions: 
\begin{eqnarray}
\label{eq:143}
&&P_{C|C}=0,\\
&&P^{PD_{1}}_{C|C}=\nonumber\\
&&-\frac{-1 + S - T + \sqrt{9 S^2 + (1 + T)^2 + 2 S (-5 + 3 T)}}{2 (-1 + S + T)},\nonumber
\end{eqnarray}
when \(P_{C|D}=0\) and: 
\begin{eqnarray}
\label{eq:48}
 &&P_{C|D}=0,\\
 &&P^{PD_{2}}_{C|D}=\nonumber\\
&&\frac{-1 + S + 3 T - \sqrt{9 - 10 S + S^2 - 14 T + 6 S T + 9 T^2}}{2 (-1 + S + T)},\nonumber
\end{eqnarray}
when \(P_{C|C}=1\). The first segment is the same as in the case of the snowdrift game. 

A population composed of players with strategies \(h_{K}=(P_{C|D},P_{C|C})\) possesses the same statistics of interactions \(\vec{\Omega}\) as a population composed of equalizer strategies \(s^{EQ}_{M_{1}}(p_{CC},p_{DD})\) if: \onecolumngrid 
\begin{eqnarray}
\label{eq:1735}
&&P_{C|D}=\\
&&\frac{(1 + p_{CC} - p_{DD}) p_{DD}}{(1 - p_{CC} + p_{DD}) (p_{DD} (1 + 2 S (-1 + T) - 2 T) + T + p_{CC} T + S (-1 + p_{CC} + 2 T - 2 p_{CC} T))},\nonumber\\
&&P_{C|C}=\nonumber\\
&&1 + \frac{(-1 + p_{CC}) (1 + p_{CC} - p_{DD})}{(-1 + p_{CC} - p_{DD}) (-1 + p_{DD} T + S (2 + p_{DD} - 2 (1 + p_{DD}) T) + p_{CC} (-1 + 2 S T))}.\nonumber
\end{eqnarray}
\twocolumngrid 
This result follows from the solution of (\ref{zd543}) for \(\vec{\Omega}\) and the definition of \(h_{K}\) strategies (\ref{eq:59}). 

Mutual information (\ref{eq:19}) shared by two identical players \(h_{K}=(P_{C|D},P_{C|C})\), following substitution of (\ref{eq:37}) to (\ref{eq:19}), is:
\onecolumngrid 
\begin{eqnarray}
\label{eq:31} I=\log\frac{(1-P_{C|D})^{(1-P_{C|D})P_{D}}P_{C|D}^{P_{C|D}P_{D}}(1-P_{C|C})^{(1-P_{C|C})(1-P_{D})}P_{C|C}^{P_{C|C}(1-P_{D})}}{P_{D}^{P_{D}}(1-P_{D})^{(1-P_{D})}},
\end{eqnarray}
\twocolumngrid 
where \(P_{D}\) is defined by (\ref{eq:24}).

\section{Appendix E: War of Attrition}
\label{sec:org22c3b93}
\label{SEC:appwaratt} 

Let us calculate the ESS of war of attrition following\cite{Smith1982,Bishop1978}. The first player wins if \(t_{1}>t_{2}\) with probability: 
\begin{eqnarray}
\label{eq:4a}
\int_{t_{1}>t_{2}}p_{1}(t_{1})p_{2}(t_{2})dt_{1}dt_{2}.
\end{eqnarray}

Thus, the payoff of the first player is: 
\begin{eqnarray}
\label{eq:7}
&&\int_{t_{1}>t_{2}}(V-g(t_{2}))p_{1}(t_{1})p_{2}(t_{2})dt_{1}dt_{2}-\\
&&\int_{t_{1}<t_{2}}g(t_{1})p_{1}(t_{1})p_{2}(t_{2})dt_{1}dt_{2},\nonumber
\end{eqnarray}

It can be rewritten as: 
\begin{eqnarray}
\label{eq:8123}
&&\int_{0}^{\infty}dt_{1}\int_{0}^{t_{1}}dt_{2}(V-g(t_{2}))p_{1}(t_{1})p_{2}(t_{2})-\\
&&\int_{0}^{\infty}dt_{1}g(t_{1})\int_{t_{1}}^{\infty}dt_{2}p_{1}(t_{1})p_{2}(t_{2}),\nonumber
\end{eqnarray}
Strategy \(p(t)\) that slightly deviates from ESS \(p_{1}=p+\delta p_{1}\) brings no advantage to the first player if \(p_{2}=p\): 
\begin{eqnarray}
\label{eq:35} 
&&\int_{0}^{\infty}dt_{1}\delta p_{1}(t1)\\
&&\left [\int_{0}^{t_{1}}dt_{2}(V-g(t_{2}))p_{2}(t_{2})-g(t_{1})\int_{t_{1}}^{\infty}dt_{2}p_{2}(t_{2})\right ]=0,\nonumber
\end{eqnarray}
for any \(\delta p_{1}\). Following (\ref{eq:35}): 
\begin{eqnarray}
\label{eq:9}
\int_{0}^{t}(V-g(t))p(t)dt-g(t)\int_{t}^{\infty}p(t)dt=0.
\end{eqnarray}
is the integral equation for the ESS strategy \(p(t)\). 

The solution of (\ref{eq:9}) is: 
\begin{eqnarray}
\label{eq:3}
p=\frac{g'}{V}\exp{-\frac{g(t)}{V}},
\end{eqnarray}
where \(p\) is the ESS probability to keep fighting at time \(t\). 

The probability of a fight to take time \(T\) is: 
\begin{eqnarray}
\label{eq:11}
P(T)=2p(T)\int_{T}^{\infty}p(t)dt,
\end{eqnarray}
Taking into account (\ref{eq:3}), the probability (\ref{eq:11}) becomes: 
\begin{eqnarray}
\label{eq:36}
P(T)=\frac{2g'}{V}\exp{\frac{-2g}{V}},
\end{eqnarray}
The probability of fight duration (\ref{eq:36}) decays exponentially with time \(t\) if the cost of fight grows linearly with time \(g(t)=Kt\). 

Sometimes, it is more convenient to work with cumulative probability of fight to take less than time \(T\): 
\begin{eqnarray}
\label{eq:32736}
\int_{0}^{T}P(t)dt=1-\exp^{-\frac{2g(T)}{V}}.
\end{eqnarray}

Both (\ref{eq:36}) and (\ref{eq:32736}) can be checked against an experiment. 


\end{document}